\documentclass[aps,prd,reprint,nofootinbib,floatfix,showkeys]{revtex4-2}
\usepackage{graphicx}
\usepackage{amsmath,amssymb,amsfonts}
\usepackage{amsthm}
\usepackage{mathrsfs}
\usepackage{capt-of}
\usepackage{xcolor}
\usepackage{booktabs}
\usepackage{mathtools}
\usepackage{placeins}
\usepackage[caption=false]{subfig}
\usepackage[
    colorlinks=true,
    linkcolor=blue,
    citecolor=blue,
    urlcolor=blue,
    pdfborder={0 0 0}
]{hyperref}

\begin{document}

    \title{Dynamical Systems Analysis of an Einstein--Cartan Ekpyrotic Nonsingular Bounce Cosmology}
    \author{Jackson Stingley}
    \affiliation{Department of Physics, Boise State University, Boise, ID, USA} 
    \affiliation{Department of Mathematics, Boise State University, Boise ID, USA}

    \begin{abstract}
        I construct an Einstein--Cartan ekpyrotic model (ECEM): a homogeneous, nearly Friedmann--Lemaître--Robertson--Walker (FLRW) background in Einstein--Cartan (EC) gravity whose spin--torsion sector, modeled phenomenologically as a Weyssenhoff fluid with stiff scaling $\rho_s\propto a^{-6}$, is coupled to a scalar field with a steep exponential potential that interpolates between a negative ekpyrotic branch and a positive plateau. Extending the Copeland--Liddle--Wands (CLW) scalar--fluid dynamical system to a six-dimensional phase space including shear, curvature, and spin--torsion, I recast the equations in a compact deceleration-parameter form, compute the full Jacobian, and evaluate maximal Lyapunov exponents. Numerical solutions show that the ekpyrotic branch ($w_\phi\gg1$) exponentially damps homogeneous shear, while the softened branch ($w_\phi<1$) allows $\rho_s$ to overtake the scalar during contraction and trigger a torsion-supported bounce at high but finite densities where the EC spin--torsion term becomes dynamically dominant. Scans in a two-parameter softening plane $(\phi_{\rm b},\Delta)$ identify a finite region of nonsingular trajectories and quantify the required tuning; in the parameter ranges explored the maximal Lyapunov exponent on the constrained phase space is negative, giving no indication of chaotic behavior in this homogeneous truncation even when the usual curvature mode that destabilizes contracting General Relativity (GR) backgrounds is included. The construction is purely phenomenological and confined to homogeneous backgrounds: it does not address entropy accumulation, the cosmological arrow of time, or a complete cyclic cosmology.
    \end{abstract}

            
    \maketitle
    \section{Introduction and motivation}
    \label{sec:intro}
    Cosmic acceleration is well measured, but its physical origin remains unclear. A cosmological constant fits current data \cite{Planck2018Parameters}, yet fine-tuning issues and present data tensions \cite{Riess2022H0,Abdalla2021H0,Abdalla2022Tensions} keep simple dynamical dark-energy models in play. Current surveys still allow small deviations from the dark-energy equation-of-state parameter $w=-1$ \cite{DESI2025PhysicsFocused}, and on large scales the universe is close to isotropic \cite{Planck2018Isotropy,Saadeh2016Isotropy}. In this paper I work with a canonical scalar field with a steep exponential potential, including a negative ekpyrotic branch, in an Einstein--Cartan (EC) setting. I use this as a simple phenomenological model of a nonsingular bounce in a homogeneous, nearly Friedmann--Lemaître--Robertson--Walker (FLRW) background, and I treat the spin--torsion sector phenomenologically as a Weyssenhoff fluid that modifies the Friedmann equations rather than deriving the full Einstein--Cartan--Sciama--Kibble (ECSK) spinor theory.

    A central motivation is to model a cosmological bounce that replaces the Big Bang singularity at the level of a homogeneous EC cosmology. Cyclic or ekpyrotic models replace the initial singularity with a bounce \cite{Khoury2001Ekpyrotic,Steinhardt2002Cyclic,Ijjas2018Bouncing}. In EC gravity the intrinsic spin of fermions sources spacetime torsion, which at very high densities acts like a negative stiff component \cite{Hehl1976Torsion,Shapiro2002Torsion,Poplawski2010Torsion}. This purely geometric effect can, in homogeneous EC models, support a nonsingular bounce without introducing new propagating degrees of freedom beyond GR plus matter.

    Cyclic scenarios, however, face both conceptual and phenomenological challenges. Conceptually, entropy must be controlled across cycles \cite{Tolman1931Entropy,Tolman1931Periodic,Ijjas2022Entropy,Matsui2024Bounce}. In this work I do not attempt to resolve entropy accumulation or derive a cosmological arrow of time; these issues lie beyond the present homogeneous model. On the observational side, proposed anomalies, such as apparently early massive galaxies or large-angle CMB features, are sometimes invoked to motivate non-standard expansion histories \cite{BoylanKolchin2023JWST,Meissner2025CCC,Shamir2022Spin,Shamir2024JWST}. In this work these connections are treated as hypotheses rather than as established facts. The aim is to analyze the internal consistency and phase-space structure of a tuned class of nonsingular expand--contract--bounce backgrounds, not to fit specific data sets or resolve particular anomalies. Accordingly, the analysis is confined to homogeneous backgrounds and their dynamical systems structure.

    The ingredients are standard: ekpyrotic contraction to suppress homogeneous shear, and EC torsion to regularize high densities. Torsion-driven bounces have already been realized in more microscopic settings. In particular, Farnsworth, Lehners, and Qiu constructed a single-field Einstein--Cartan model in which a spinor condensate drives both ekpyrotic contraction and a torsion-induced bounce, and analyzed the associated cosmological perturbations in detail \cite{Farnsworth2017SpinorBounce}. In their construction the spinor bilinear plays the role of a scalar order parameter, and the focus is on exact background solutions and their perturbations rather than on the global phase-space structure of a more general homogeneous system.

    In contrast, the present work adopts a two-component description: a canonical scalar field provides the ekpyrotic dynamics, while a Weyssenhoff-type spin fluid provides an effective phenomenological description of the torsion sector. Keeping these as distinct components makes it straightforward to set up a global dynamical-systems analysis of the scalar--fluid--spin system, track fixed points and basins of attraction, and study stability of the background evolution. I extend the Copeland--Liddle--Wands (CLW) formalism to a six-dimensional phase space $(x,y,z,\Sigma,\Omega_k,\Omega_s)$ and recast the equations in a compact deceleration-parameter $q$-form. Within this framework I construct a tuned expand--contract--bounce \emph{scenario}. A single scalar field with a softening potential produces scalar-dominated expansion, a fluid-dominated era, ekpyrotic contraction, and a torsion-regulated bounce in one homogeneous setup. The contraction-to-expansion bounce is demonstrated by direct cosmic-time integration in a spatially flat benchmark run ($k=0$, Fig.~\ref{fig:bounce_results}), and a separate positively curved run ($k=+1$, Fig.~\ref{fig:ecem_cycle}) shows repeated turnaround and bounce crossings in the same homogeneous system. The softening is chosen so that the ekpyrotic phase transitions to $w_\phi<1$ just as the spin--torsion term becomes dynamically important; this is a controlled but non-generic choice, treated here as a phenomenological tuning rather than a prediction. In this tuned setup I show that a steep negative branch of the potential produces an ekpyrotic phase that damps homogeneous shear, and that for trajectories in which the spin--torsion sector already dominates the stiff component, the Weyssenhoff fluid can overtake curvature and trigger a torsion-supported bounce below the Planck scale. Throughout, I focus on how the existence of a nonsingular ekpyrotic--bounce background depends on the softening parameters and on the qualitative phase-space structure of the homogeneous dynamical system. Explicit data-driven constraints and a full treatment of inhomogeneous perturbations are left for future work.

    \subsection{Related work and novelty}
    \label{sec:novelty}
    Scalar-field plus fluid systems with exponential potentials have been studied extensively in GR \cite{Copeland1998Scaling,Heard2002,WainwrightEllis1997}. Billyard et al.\ showed that the CLW scaling solution is stable to shear in expanding GR backgrounds \cite{Billyard1998Scaling}, while in contracting GR backgrounds a canonical scalar is generically unstable to anisotropy (the shear instability), which motivates ekpyrotic phases with $w_\phi \gg 1$ \cite{Erickson2004Ekpyrotic}. Torsion-induced bounces have also been realized in Einstein--Cartan and related torsionful models, including spinor-based constructions in which a single Dirac field drives both ekpyrotic contraction and a torsion-supported bounce \cite{Hehl1976Torsion,Shapiro2002Torsion,Poplawski2010Torsion,Farnsworth2017SpinorBounce,Poplawski2020,Alam2025}. As noted above, most of these papers focus on the spin-sector microphysics and on particular background or perturbative solutions. For example, Ref.~\cite{Farnsworth2017SpinorBounce} studies exact spinor-driven ekpyrotic bounce backgrounds and their perturbations, rather than the global phase-space structure of a scalar--fluid--spin system in a dynamical-systems picture.

    The present work does \emph{not} introduce a new microscopic bounce mechanism. The novelty lies at the level of the homogeneous dynamical system:
    (i) I cast an Einstein--Cartan cosmology with a canonical scalar, barotropic fluid, homogeneous shear, curvature, and a Weyssenhoff spin fluid into a six-dimensional Copeland--Liddle--Wands--type phase space $(x,y,z,\Sigma,\Omega_k,\Omega_s)$ obeying the modified Friedmann constraint \eqref{eq:Friedmann}.
    (ii) I rewrite the system in a compact deceleration-parameter $q$-form \eqref{eq:q_canonical} and derive the full $6{\times}6$ Jacobian and eigenvalue spectrum (Secs.~\ref{sec:eigenvalues}, \ref{app:ecem_jacobian}), making the coupling between the scalar--fluid sector and the geometric sector $(\Sigma,\Omega_k,\Omega_s)$ explicit.
    (iii) Within this framework, I show how a \emph{single} scalar with a softening potential can realize scalar-dominated expansion, a matter/radiation era, ekpyrotic contraction with homogeneous shear suppression, and a torsion-induced bounce within one consistent homogeneous setup. By scanning a two-parameter softening plane $(\phi_{\rm b},\Delta)$ I identify a finite region of parameter space that yields nonsingular ekpyrotic--bounce trajectories, providing an explicit ``bounce-basin'' characterization of the required tuning (Sec.~\ref{sec:bounce-basin}, Fig.~\ref{fig:bounce_basin}).
    (iv) I supplement the linear stability analysis with a maximal Lyapunov exponent computation on the constrained phase space, finding $\lambda_{\max}<0$ for the ranges tested and hence, within this homogeneous truncation, no indication of chaotic behavior over the parameter ranges explored (Sec.~\ref{sec:lyapunov}).
    To the best of my knowledge, a global dynamical-systems analysis of an Einstein--Cartan cosmology with a canonical scalar, barotropic fluid, shear, curvature, and an effective Weyssenhoff spin fluid, with explicit bounce-basin and Lyapunov characterization, has not been carried out before. In this way the construction is complementary to microphysics-focused ECSK spinor models and to spinor-driven ekpyrotic bounce scenarios such as Ref.~\cite{Farnsworth2017SpinorBounce}: the emphasis here is on the global phase-space structure of a two-component (scalar + spin fluid) system, mapping fixed points, basins of attraction, and Lyapunov stability, while treating parameters such as the Weyssenhoff coefficient $\alpha$ and the softening scale as model inputs rather than deriving them from a specific ultraviolet completion.

    \section{Assumptions and conventions}
    \label{sec:assumptions}
    I work in units $c=\hbar=8\pi G=1$. The time variable is the number of e-folds,
    \begin{equation*}
        N \equiv \ln a.
    \end{equation*}
    A prime denotes differentiation with respect to $N$ and is related to cosmic time $t$ via
    \begin{equation*}
        \frac{d}{dN} = \frac{1}{H}\frac{d}{dt},
    \end{equation*}
    and overdots denote derivatives with respect to $t$. I assume a nearly FLRW background with small homogeneous shear \cite{WainwrightEllis1997,Chan2012Thesis}. Homogeneous anisotropy is modeled by a single shear scalar $\sigma$; I use the normalized variable $\Sigma \equiv \sigma/H$ in the dynamical system, corresponding to a Bianchi~I-type degree of freedom. I do not include full Bianchi~IX dynamics or inhomogeneous Belinski--Khalatnikov--Lifshitz (BKL) behavior.

    \begin{table}[ht]
        \centering
        {\tiny
        \begin{tabular}{|c|l|}
            \hline
            \textbf{Symbol} & \textbf{Definition / Physical Meaning} \\
            \hline
            $t, N$ & Cosmic time and e-fold time ($N \equiv \ln a$) \\
            $a, H$ & Scale factor and Hubble parameter ($H \equiv \dot{a}/a$) \\
            $k$ & Spatial curvature index ($k = +1,0,-1$) \\
            $q$ & Deceleration parameter ($q \equiv -1 - \dot{H}/H^2$) \\
            \hline
            $\phi, V(\phi)$ & Canonical scalar field and potential \\
            $x, y$ & Normalized kinetic and (absolute) potential energy of $\phi$ \\
            $w_\phi$ & Scalar field equation of state ($w_\phi \equiv p_\phi/\rho_\phi$) \\
            \hline
            $\rho_m$ & Background matter energy density \\
            $\rho_s$ & Spin--torsion energy density ($\rho_s \propto a^{-6}$) \\
            $\rho_{\rm tot}, p_{\rm tot}$ & Total energy density and pressure \\
            $w_m, \gamma_m$ & Matter equation of state and barotropic index ($\gamma_m \equiv 1 + w_m$) \\
            $w_{\rm tot}$ & Total equation of state ($w_{\rm tot} \equiv p_{\rm tot}/\rho_{\rm tot}$) \\
            \hline
            $z$ & Normalized background matter density ($z \equiv \rho_m / (3H^2)$) \\
            $\Sigma, \sigma$ & Normalized shear and shear scalar ($\Sigma \equiv \sigma/H$) \\
            $\Omega_k$ & Spatial curvature parameter ($\Omega_k \equiv -k/(a^2 H^2)$) \\
            $\Omega_s$ & Spin--torsion density parameter ($\Omega_s \equiv \rho_s/(3H^2)$) \\
            \hline
            $\lambda(\phi)$ & Dynamical slope of the potential ($\lambda \equiv -V_{,\phi}/V$) \\
            $s$ & Sign of the scalar potential ($s \equiv \mathrm{sign}(V) = \pm 1$) \\
            \hline
        \end{tabular}
        }
        \caption{Summary of mathematical notation and physical variables used in the model.}
        \label{tab:symbols}
    \end{table}
    \FloatBarrier

    Spatial curvature is $k\in\{+1,0,-1\}$, with
    \begin{equation}
        \Omega_k \equiv -\frac{k}{a^2 H^2}
        \quad\Rightarrow\quad
        k=+1 \ \Longleftrightarrow\ \Omega_k<0.
    \end{equation}

    To suppress anisotropic shear during contraction while still allowing a nonsingular bounce, I use a \textbf{canonical} scalar field with Lagrangian
    \begin{equation}
        \mathcal{L} = X - V(\phi), \qquad
        X \equiv \frac{1}{2}\dot{\phi}^2.
    \end{equation}
    I introduce the usual normalized variables for a canonical scalar,
    \begin{equation}
        x \equiv \frac{\dot{\phi}}{\sqrt{6}\,H}, \qquad
        y \equiv \frac{\sqrt{|V|}}{\sqrt{3}\,H},
    \end{equation}
    in terms of which the scalar equation of state is
    \begin{equation}
    w_\phi(N) = \frac{x^2 - s y^2}{x^2 + s y^2},
    \end{equation}
    where $s \equiv \mathrm{sign}(V) \in \{+1,-1\}$ tracks the sign of the potential.

    The matter fluid has constant equation of state $w_m$, and I define
    \begin{equation}
        \gamma_m \equiv 1+w_m
        \qquad
        (\text{dust: }\gamma_m=1,\ \text{radiation: }\gamma_m=\tfrac{4}{3}).
    \end{equation}
    Together with the scalar variables $(x,y)$, I extend the dynamical-system variables to include the matter, shear, and spin sectors via
    \begin{align}
        z &= \frac{\rho_m}{3H^2}, \qquad
        \Sigma = \frac{\sigma}{H}, \qquad
        \Omega_s = \frac{\rho_s}{3H^2},
    \end{align}
    where the homogeneous shear contributes an effective stiff component with energy density
    \begin{equation}
        \rho_\sigma = 3H^2 \Sigma^2, \qquad w_\sigma = 1.
    \end{equation}
    In these variables the normalized quantities $(x,y,z,\Sigma,\Omega_k,\Omega_s)$ satisfy the modified Friedmann constraint
    \begin{equation}
        \label{eq:Friedmann}
        x^2 + s\,y^2 + z + \Sigma^2 + \Omega_k - \Omega_s = 1,
    \end{equation}
    where $\rho_s>0$ is defined such that $\Omega_s>0$ corresponds to a repulsive contribution to the Friedmann equation.

    I adopt a \textit{softening potential} that transitions from a steep negative exponential (driving ekpyrosis, $w_\phi\gg1$) to a positive plateau. The positive plateau is important because it allows a phase with $w_\phi<1$, changing the redshifting of the scalar energy density,
    \begin{equation}
        \rho_\phi \propto a^{-3(1+w_\phi)},
    \end{equation}
    relative to the spin--torsion density $\rho_s\propto a^{-6}$. If $w_\phi<1$, then $\rho_\phi$ scales more slowly than $\rho_s$, so in a contracting universe ($a\to0$) the repulsive spin term can eventually overtake the scalar and enable a bounce. A concrete monotone realization is
    \begin{equation}
        \label{eq:scalar_potential}
        V(\phi) = -V_0 e^{-\lambda_0 \phi}\,\bigl[1 - \mathcal{S}(\phi)\bigr]
                + V_{\rm soft}\,\mathcal{S}(\phi),
    \end{equation}
    with $V_0>0$, $\lambda_0^2>6$ setting the ekpyrotic steepness, and $V_{\rm soft}>0$ the positive plateau value. The switching function
    \begin{equation}
        \mathcal{S}(\phi) \equiv \frac{1}{2}\left[1 + \tanh\!\left(\frac{\phi-\phi_{\rm b}}{\Delta}\right)\right]
    \end{equation}
    mediates the transition around $\phi_{\rm b}$ with width $\Delta>0$. For $\phi\ll\phi_{\rm b}$ one has $\mathcal{S}\to0$ and $V\simeq -V_0 e^{-\lambda_0\phi}$ (the ekpyrotic regime); for $\phi\gg\phi_{\rm b}$ one has $\mathcal{S}\to1$ and $V\to V_{\rm soft}>0$, creating a regime with $w_\phi<1$ in which the EC spin--torsion sector can dominate and produce a bounce.

    In this work I treat $V(\phi)$ as an effective potential rather than a fundamental microphysical model, in the same spirit as standard dark-energy parameterizations \cite{Chevallier2001,Linder2003}. The transition scale set by $(\phi_{\rm b},\Delta)$ is chosen so that the softening of the ekpyrotic branch happens in the same \emph{order-of-magnitude} density range in which the EC spin--torsion term would otherwise enforce a bounce. In a UV-complete theory it is natural for both effects to be tied to a single characteristic mass scale, but not necessarily at exactly the same density. Here I simply tune them to overlap: the parameters $(\phi_{\rm b},\Delta,V_{\rm soft})$ are chosen so that the transition to a regime with $w_\phi<1$ coincides with the range in which the EC spin--torsion term becomes dynamically dominant. The numerical scan in Sec.~\ref{sec:bounce-basin} (Fig.~\ref{fig:bounce_basin}) shows that the bounce appears over a finite $\mathcal{O}(1)$ region of this parameter space rather than at a single point. This one-scale parametrization should be read as a working assumption about the high-density completion: it is tuned to the bounce scale, but not point-fine-tuned to an isolated delta in parameter space.

    \section{Einstein--Cartan field equations}
    \label{sec:EC-field-eqs}
    The coupling of spin to torsion is described by the Einstein--Cartan--Sciama--Kibble action, where the affine connection has an antisymmetric part and the spin current sources torsion~\cite{Hehl1976Torsion,Shapiro2002Torsion}. Varying the action with respect to the torsion tensor $S^{\lambda}{}_{\mu\nu}$ gives an algebraic relation between torsion and the spin density. For a Dirac field $\psi$:
    \begin{equation}
        S^{\lambda}{}_{\mu\nu} \propto \varepsilon^{\lambda}{}_{\mu\nu\rho}\,
        \bar{\psi}\,\gamma^{\rho}\gamma^{5}\,\psi,
    \end{equation}
    so torsion does not propagate; it is fixed locally by the axial spin current. Substituting this relation back into the action generates a four-fermion self-interaction for the Dirac field, and the corresponding correction to the energy density scales as
    \begin{equation}
        \delta\rho_{\rm tor} \sim S^2 \sim
        \bigl(\bar{\psi}\gamma^\mu\gamma^5\psi\bigr)
        \bigl(\bar{\psi}\gamma_\mu\gamma^5\psi\bigr)
        \sim n^2,
    \end{equation}
    where $n$ is the fermion number density~\cite{Weyssenhoff1947,Hehl1976Torsion,Poplawski2010Torsion}. For a conserved fermion current one has $n \propto a^{-3}$ in an expanding or contracting FLRW background, so the spin--torsion contribution behaves as a stiff component,
    \begin{equation}
        \rho_s \propto n^2 \propto a^{-6}.
    \end{equation}
    As a result, I model the spin--torsion correction as a stiff component with
    \begin{equation}
        \rho_s \equiv \alpha a^{-6}, \qquad \alpha>0.
    \end{equation}
    In Einstein--Cartan theory, algebraically eliminating torsion gives a quadratic spin correction. With the conventions used here, that correction enters the effective FLRW background equations with negative sign. I denote its magnitude by $\rho_s>0$ and write the torsion sector as an effective fluid,
    \[
        \rho_{\rm tor} \equiv -\rho_s,\qquad p_{\rm tor} \equiv -\rho_s,
    \]
    so it has stiff equation of state ($w_{\rm tor}=1$) while contributing negative effective density and pressure. The corresponding effective background source is
    \[
        \rho_{\rm eff}=\rho_{\rm tot}+\rho_{\rm tor}=\rho_{\rm tot}-\rho_s,\qquad
        p_{\rm eff}=p_{\rm tot}+p_{\rm tor}=p_{\rm tot}-\rho_s,
    \]
    and
    \[
        \rho_{\rm eff}+p_{\rm eff}=\rho_{\rm tot}+p_{\rm tot}-2\rho_s.
    \]
    So the $-\rho_s$ term in Friedmann and the $-2\rho_s$ term in Raychaudhuri follow directly from the algebraic Einstein--Cartan/Weyssenhoff reduction, rather than from an added sign choice~\cite{Hehl1976Torsion,Shapiro2002Torsion,Poplawski2010Torsion}. For nonrelativistic matter, $\rho_m \simeq mn$, so the microscopic Weyssenhoff scaling gives $\rho_s\propto n^2\propto \rho_m^2$ up to constant factors. In the phenomenological dynamical-systems treatment below, $\Omega_s$ and $z$ are kept independent.

    The EC-modified background equations can be written as
    \begin{equation}
        \label{eq:EC-Friedmann}
        \begin{aligned}
            H^2 &= \tfrac{1}{3}\big(\rho_{\rm tot} - \rho_s\big) - \tfrac{k}{a^2},\qquad
            \dot{H} &= -\tfrac{1}{2}\big(\rho_{\rm tot} + p_{\rm tot} - 2\rho_s\big) + \tfrac{k}{a^2},
        \end{aligned}
    \end{equation}
    where $\rho_{\rm tot}$ and $p_{\rm tot}$ are the combined energy density and pressure of the scalar field, matter, and shear sector only (excluding the spin--torsion contribution $\rho_s$, which is treated separately). The associated density parameter is
    \begin{equation}
        \Omega_s = \frac{\rho_s}{3H^2} = \frac{\alpha a^{-6}}{3H^2},
    \end{equation}
    so that the $-\Omega_s$ term in \eqref{eq:Friedmann} encodes the repulsive $-\rho_s$ contribution in \eqref{eq:EC-Friedmann}.

    For a spatially flat background ($k=0$), a nonsingular bounce is defined by $H=0$ and $\dot H>0$. Here $\rho_{\rm tot}$ and $p_{\rm tot}$ contain the scalar, fluid, and shear sectors but \emph{exclude} the spin--torsion piece, which is entirely encoded in $\rho_s$. At $H=0$, Eq.~\eqref{eq:EC-Friedmann} implies
    \begin{equation}
        \rho_{\rm tot} = \rho_s,
    \end{equation}
    while the Raychaudhuri equation gives
    \begin{equation}
        \dot H = -\tfrac{1}{2}\bigl(\rho_{\rm tot} + p_{\rm tot} - 2\rho_s\bigr)
            = \tfrac{1}{2}\bigl(\rho_{\rm tot} - p_{\rm tot}\bigr)
            = \tfrac{1}{2}\rho_{\rm tot}\,(1 - w_{\rm tot}),
    \end{equation}
    with $w_{\rm tot} \equiv p_{\rm tot}/\rho_{\rm tot}$. Since the shear sector has $w_\sigma = 1$, any component with $w_{\rm tot}<1$ must be supplied by the scalar and/or fluid. Thus $\dot H>0$ at the bounce requires
    \begin{equation}
        w_{\rm tot} < 1,
    \end{equation}
    so equality of energy densities is necessary but not sufficient: the total equation of state has to soften below the stiff bound $w=1$. If the ekpyrotic phase with $w_\phi \gg 1$ were to persist all the way to the equality point $\rho_{\rm tot} = \rho_s$, one would have $\dot H<0$ there and the EC spin--torsion term would simply delay, rather than remove, a curvature singularity. In the model studied here the softening branch of the potential is a kinematic requirement: it is chosen so that $w_\phi<1$ near the would-be equality scale, driving $w_{\rm tot}<1$ and allowing the spin--torsion term to generate a nonsingular bounce. For $k=+1$ the curvature term tightens the bounce condition; $w_{\rm tot}<1$ remains necessary but is not in general sufficient.

    \section{Extended dynamical system}
    Using e-fold time $N = \ln a$, I define the dynamical slope of the potential as
    \begin{equation}
        \lambda(\phi) \equiv -\frac{V_{,\phi}}{V}.
    \end{equation}
    Because the softened potential \eqref{eq:scalar_potential} interpolates between a negative exponential branch and a positive plateau, it necessarily passes through a zero at some $\phi = \phi_0$ with $V(\phi_0) = 0$. At that point $\lambda(\phi)$ diverges, but the combination $s\,\lambda y^2$ that enters the $x'$ equation remains finite, since
    \[
        s\,\lambda y^2 = -\frac{V_{,\phi}}{3H^2}
    \]
    is regular. Near the zero the $y$-equation in $(x,y)$ variables becomes ill-conditioned, so in numerical work I instead evolve $\phi$ (and hence $V(\phi)$ and $V_{,\phi}$) directly across the zero rather than $\lambda(\phi)$ itself.
    The coupling between the scalar--fluid sector and the geometric variables is mediated by the deceleration parameter $q$. Using the EC-modified Raychaudhuri equation including spatial curvature, $q$ can be written in normalized variables as
    \begin{equation}
        \label{eq:q_canonical}
        q = 3x^2 - 1 + \frac{3}{2}\gamma_m z + 3\Sigma^2 + \Omega_k - 3\Omega_s.
    \end{equation}
    This form makes the coupling between the geometric sector $(\Sigma,\Omega_k,\Omega_s)$ and the scalar--fluid sector $(x,y,z)$ explicit.

    In terms of the variables $(x,y,z,\Sigma,\Omega_k,\Omega_s)$, the evolution equations can be written as
    \begin{align}
        x'        &= -3x + s\sqrt{\tfrac{3}{2}}\,\lambda y^2 + x(1+q), &
        y'        &= -\sqrt{\tfrac{3}{2}}\,\lambda xy + y(1+q), \label{eq:dyn_xy}\\
        z'        &= z\bigl[-3\gamma_m + 2(1+q)\bigr],                &
        \Sigma'   &= (q-2)\Sigma,                                    \label{eq:dyn_zsigma}\\
        \Omega_k' &= 2q\,\Omega_k,                                   &
        \Omega_s' &= 2(q-2)\Omega_s.                                 \label{eq:dyn_ks}
    \end{align}
    For a pure exponential potential with constant $\lambda$ the system \eqref{eq:dyn_xy}-\eqref{eq:dyn_ks} is autonomous in $(x,y,z,\Sigma,\Omega_k,\Omega_s)$ and admits the familiar CLW fixed points. For the softened potential \eqref{eq:scalar_potential} I treat $\lambda(\phi)$ as a slowly varying background function and analyze fixed points only in the asymptotic regimes where $\lambda \approx \text{const}$, while the full evolution between those regimes is obtained numerically.

    \subsection{Shear suppression}
    \label{sec:shear_suppression}
    In the contracting phase driven by the steep negative branch of the potential, the ekpyrotic scalar dominates and $w_\phi \gg 1$. Neglecting subdominant components, the deceleration parameter is approximately
    \begin{equation}
        q \simeq \frac{1}{2}\bigl(1+3w_\phi\bigr) \gg 2,
    \end{equation}
    and varies slowly over the ekpyrotic regime. The shear evolution equation
    \begin{equation}
        \Sigma' = (q-2)\Sigma
    \end{equation}
    then integrates to
    \begin{equation}
        \Sigma(N) \propto \exp\!\Big[\int (q-2)\,dN\Big]
        \simeq \exp[(q-2)N] = a^{\,q-2},
    \end{equation}
    where the last step assumes $q$ approximately constant during the ekpyrotic phase. Since the universe is contracting ($N \to -\infty$) and $q-2>0$, the shear $\Sigma$ decays exponentially as the universe evolves toward the bounce.

    The same conclusion follows from the energy-density scalings. The shear energy density scales as $\rho_\sigma \propto a^{-6}$, while the ekpyrotic scalar scales as
    \begin{equation}
        \rho_\phi \propto a^{-3(1+w_\phi)}, \qquad w_\phi>1,
    \end{equation}
    so that, for the ekpyrotic equation of state $w_\phi>1$,
    \begin{equation}
        \frac{\rho_\sigma}{\rho_\phi} \propto a^{3(1+w_\phi)-6} \;\longrightarrow\; 0
        \quad\text{as}\quad a\to 0.
    \end{equation}
    Thus the anisotropic contribution becomes negligible during ekpyrotic contraction, and at the homogeneous level the ekpyrotic phase acts as a strong attractor towards isotropy, suppressing the growth of shear anisotropy.\footnote{Strictly speaking, the Belinski--Khalatnikov--Lifshitz (BKL) analysis concerns the growth of spatial gradients and mixmaster-type behavior in generic inhomogeneous collapse. The present work addresses the homogeneous (Bianchi-type) limit, in which controlling the shear sector is a necessary but not sufficient condition for taming the full BKL dynamics.} This matches the behavior found in Refs.~\cite{Erickson2004Ekpyrotic,Garfinkle2008Ekpyrotic}, before the system enters the softening regime where the EC spin--torsion sector becomes dynamically relevant.

    \section{Phase-space structure of ECEM evolution}
    \label{sec:phase-space-structure}
    In the constant-slope regimes where $\lambda \approx \text{const}$, the system admits a set of fixed points and asymptotic dynamical regimes. For present purposes it is convenient to group the relevant behaviors into four broad classes:

    \begin{enumerate}
        \item \textbf{Scalar-field dominated (possibly accelerated).}
        Here the scalar carries essentially all of the energy density, with $(x^2 + s y^2) \simeq 1$ and $\Sigma \simeq 0 \simeq \Omega_k \simeq \Omega_s$; on the positive-potential branch ($s=+1$) this reduces to $(x^2 + y^2) \simeq 1$. For constant slope $\lambda$ there is a scalar-dominated fixed point whenever $\lambda^2<6$; it drives accelerated expansion for $\lambda^2<2$ \cite{Copeland1998Scaling} and is stable against matter perturbations if $\lambda^2<3\gamma_m$ \cite{Copeland1998Scaling}. In the extended Einstein--Cartan ekpyrotic model (ECEM) phase space, $\lambda^2<6$ ensures that anisotropy and spin--torsion perturbations decay, while curvature perturbations decay only in the accelerated regime $\lambda^2<2$ (see Sec.~\ref{sec:eigenvalues}).
        
        \item \textbf{Fluid dominated (matter or radiation).}
        Here $x \simeq 0$, $y \simeq 0$, $z \simeq 1$, corresponding to the usual radiation and matter dominated eras \cite{WainwrightEllis1997}. Their full eigenvalue spectrum in the $(x,y,z,\Sigma,\Omega_k,\Omega_s)$ system is given in Sec.~\ref{sec:eigenvalues} and Table~\ref{tab:extended_fp}; for $0<\gamma_m\le 2$ they appear as saddle points.
        
        \item \textbf{Scaling (tracking).}
        For $\lambda^2>3\gamma_m$, the scalar tracks the fluid with $w_\phi = w_m$ and $\Omega_\phi = 3\gamma_m/\lambda^2$ \cite{Copeland1998Scaling,Heard2002}. This point is stable in the $(x,y)$ sector but, once curvature is included, the curvature direction is unstable for $\gamma_m\ge 1$ (the GR result of \cite{Billyard1998Scaling}). In the full six-dimensional space the scaling point is a saddle.
        
        \item \textbf{Ekpyrotic contraction (steep potential regime).}
        On the contracting branch ($H<0$) the field rolls down the steep negative part of the potential ($V<0$) into an ekpyrotic phase with $w_\phi\gg 1$. As shown in Sec.~\ref{sec:shear_suppression}, this super-stiff phase drives the homogeneous shear fraction $\Sigma$ to zero exponentially, so the isotropic solution is a strong attractor \cite{Erickson2004Ekpyrotic,Garfinkle2008Ekpyrotic}. This is in sharp contrast with the kinetic case ($w_\phi=1$), where shear is only marginally stable. Contraction ends when the potential softens so that $w_\phi<1$ for a finite interval, allowing the spin--torsion density $\Omega_s\propto a^{-6}$ to overtake the scalar density and trigger a nonsingular bounce, for softening parameters in the finite bounce basin identified in Sec.~\ref{sec:bounce-basin} (Fig.~\ref{fig:bounce_basin}).
    \end{enumerate}

\begin{figure}[htpb!]
        \centering
        \includegraphics[width=.6\linewidth]{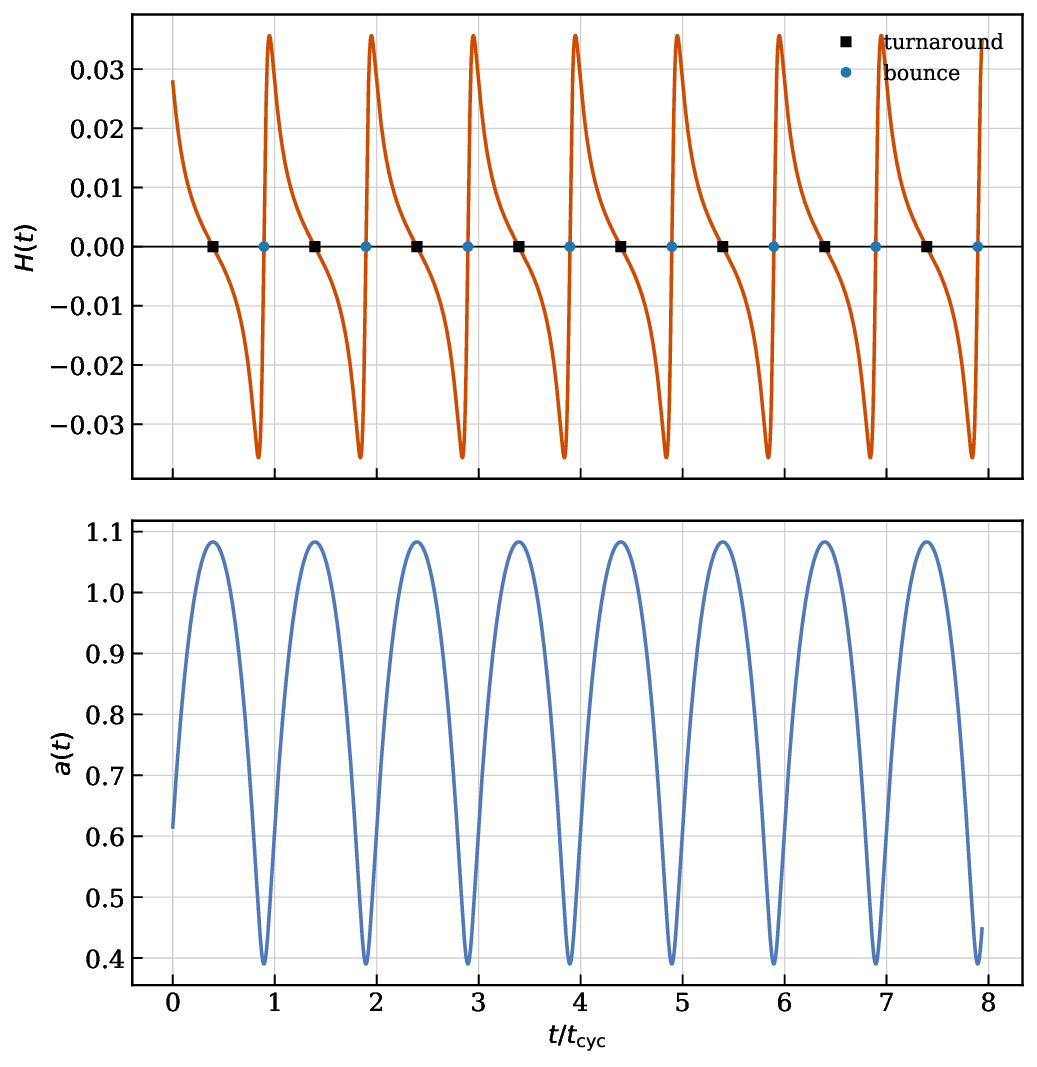}
        \caption{Numerically integrated positively curved ECEM background over multiple cycles (representative $k=+1$ run). Top: Hubble parameter $H$ versus $t/t_{\rm cyc}$, with alternating turnarounds ($H=0,\dot H<0$) and torsion-induced bounces ($H=0,\dot H>0$). Bottom: scale factor $a$ versus $t/t_{\rm cyc}$ over the same interval. This provides an explicit homogeneous expansion$\to$contraction$\to$bounce trajectory in cosmic time; in this example $a(t)$ is bounded and oscillatory rather than showing net cycle-to-cycle growth. By contrast, Fig.~\ref{fig:bounce_results} is the spatially flat ($k=0$) single-bounce benchmark used to isolate the torsion-regulated bounce itself.}
        \label{fig:ecem_cycle}
    \end{figure}

    The bounce itself is not a fixed point; it occurs when the spin density reaches the EC scale at which
    $\rho_s \simeq \rho_{\rm tot} \equiv \rho_{\rm b}$ and, provided $w_{\rm tot}<1$, the Einstein--Cartan Friedmann and Raychaudhuri equations allow $H$ to pass through zero with $\dot H>0$. At the background level this is a nonsingular $\rho^2$-type bounce, similar in spirit to the effective Friedmann equation in loop quantum cosmology \cite{AshtekarSingh2011} and to other torsion-based bouncing models \cite{Farnsworth2017SpinorBounce,Alam2025}. Trajectories in the bounce basin typically follow the sequence
    \[
    \begin{gathered}
        \text{scalar-dominated expansion} \;\to\; \text{fluid era}\\
        \to\; \text{ekpyrotic contraction}\\
        \text{(with decaying shear)}\\
        \to\; \text{torsion-driven bounce} \;\to\; \text{renewed expansion}.
    \end{gathered}
    \]
    as shown numerically in Fig.~\ref{fig:ecem_cycle} for a representative positively curved trajectory. Section~\ref{sec:bounce-basin} and Fig.~\ref{fig:bounce_basin} make precise which softening parameters $(\phi_{\rm b},\Delta)$ lead to a torsion-induced bounce rather than a singular crunch in the high-density contracting leg.

    \section{Fixed points and linear stability}
    \label{sec:eigenvalues}
    Linear stability is obtained by evaluating the Jacobian of the six-dimensional autonomous system at each critical point, subject to the Friedmann constraint. I follow the Copeland--Liddle--Wands (CLW) procedure \cite{Copeland1998Scaling}, in the spirit of standard dynamical-systems treatments of cosmology \cite{WainwrightEllis1997,Chan2012Thesis}, and extend the Jacobian to include the geometric variables $(\Sigma,\Omega_k,\Omega_s)$.

    For the fixed-point analysis I treat $\lambda$ as constant, corresponding either to a pure exponential potential or to the local value $\lambda_* \equiv \lambda(\phi_*)$ at each critical point. In the full ECEM potential~\eqref{eq:scalar_potential}, the slope $\lambda(\phi)\equiv -V_{,\phi}/V$ varies smoothly between the steep ekpyrotic branch and the softened plateau, with the transition localized over a width $\Delta$ in field space. The fixed points and eigenvalues below should be read as describing the instantaneous attractors of the system in the two adiabatic regimes where the potential is approximately exponential, rather than exact global attractors of the variable $\lambda(\phi)$ dynamics.

    \subsection{Kinetic-dominated points}
    There are two kinetic-dominated solutions corresponding to a stiff fluid ($w_\phi = 1$):
    \begin{equation}
        (x,y,z,\Sigma,\Omega_k,\Omega_s) = (\pm 1,\,0,\,0,\,0,\,0,\,0).
    \end{equation}
    At these points $q=2$. Substituting $w_\phi=1$ and $q=2$ into the general Jacobian gives the eigenvalues in Table~\ref{tab:extended_fp}. Two eigenvalues vanish, associated with shear and spin--torsion ($\mu_\Sigma = 0$ and $\mu_s = 0$). A purely stiff component neither damps nor amplifies anisotropy relative to the background ($\Sigma'=0$) and does not by itself solve the anisotropy problem of contracting GR backgrounds; addressing that requires an ekpyrotic phase with $w_\phi>1$.

    The stability convention is as follows. Since $d/dN = (1/H)\,d/dt$, the e-fold variable $N$ increases toward $+\infty$ along the expanding branch ($H>0$) and decreases toward $-\infty$ along the contracting branch ($H<0$). I define "stable" to mean decay as physical time $t\to +\infty$. For a linear mode $X'=\mu X$ this implies
    \begin{equation}
        X(N) \propto e^{\mu (N-N_0)}.
    \end{equation}
    On the expanding branch stability corresponds to $\mu<0$ (decay as $N\to +\infty$), whereas on the contracting branch stability corresponds to $\mu>0$ (decay as $N\to -\infty$). All eigenvalues quoted below are eigenvalues of the Jacobian with respect to $N$, and these sign rules should be used when interpreting stability on the expanding versus contracting branches.

    \subsection{Stability of the ekpyrotic regime}
    The ekpyrotic contraction is a dynamical regime rather than an isolated fixed point. When the scalar dominates, $q \simeq (1+3w_\phi)/2$ and the homogeneous shear obeys $\Sigma'=(q-2)\Sigma$. In the fixed-point language it is convenient to encode this as an effective shear eigenvalue
    \begin{equation}
        \mu_\Sigma^{\rm eff} \simeq \frac{3}{2}\bigl(w_\phi - 1\bigr),
    \end{equation}
    which is positive for $w_\phi>1$. Along the contracting branch $N\to -\infty$, a mode with $\Sigma'=\mu_\Sigma^{\rm eff}\Sigma$ decays as
    \begin{equation}
        \Sigma(N) \propto e^{\mu_\Sigma^{\rm eff} N} \xrightarrow{N\to -\infty} 0,
    \end{equation}
    so the ekpyrotic regime is a strong homogeneous attractor towards isotropy \cite{Erickson2004Ekpyrotic}, unlike the marginally stable kinetic points with $w_\phi=1$.

    \subsection{Fluid-dominated point}
    In addition to the scalar fixed points, the extended system has the usual barotropic fluid point
    \begin{equation}
        (x^*,y^*,z^*,\Sigma^*,\Omega_k^*,\Omega_s^*)
        = (0,0,1,0,0,0),
    \end{equation}
    corresponding to a pure fluid with $w_m = \gamma_m - 1$. The deceleration parameter is then
    \begin{equation}
        q = \frac{1}{2}\bigl(1+3w_m\bigr)
        = \frac{3\gamma_m - 2}{2}.
    \end{equation}
    Evaluating the general Jacobian of Appendix~\ref{app:ecem_jacobian} at this point yields the eigenvalues in Table~\ref{tab:extended_fp}. For standard fluids $0<\gamma_m\le 2$ this point is not a late-time attractor on the expanding branch: at least one of the scalar or curvature directions has a positive eigenvalue (for dust or radiation, $\mu_k>0$), so the fluid point is a saddle in the full six-dimensional phase space.

    \subsection{Scalar-field dominated point}
    On the positive-potential branch $s=+1$ and for $\lambda^2<6$ there is a scalar-dominated critical point
    \begin{equation}
        \Big(x^*,y^*,z^*,\Sigma^*,\Omega_k^*,\Omega_s^*\Big)
        =
        \Big(\tfrac{\lambda}{\sqrt{6}},\ \sqrt{1-\tfrac{\lambda^2}{6}},\ 0,\ 0,\ 0,\ 0\Big),
    \end{equation}
    with Jacobian spectrum given in Table~\ref{tab:extended_fp}. One eigenvalue, $\mu_\perp=\lambda^2$, corresponds to motion orthogonal to the Friedmann-constraint surface and is removed once the constraint is imposed; the physical stability is governed by the remaining five eigenvalues tangent to the constraint manifold. On the expanding branch, the conditions $\lambda^2<3\gamma_m$ and $\lambda^2<2$ (together with the existence condition $\lambda^2<6$) ensure that all geometric and matter perturbations decay, so the point is a late-time attractor. These reproduce the usual CLW criteria for accelerated expansion and stability against fluid perturbations, now extended to include curvature and torsion.

    \subsection{Scaling (tracking) point}
    On the positive-potential branch $s=+1$, when $\lambda^2>3\gamma_m$ there is a scaling solution in which the scalar tracks the background fluid:
    \begin{equation}
        \left(
            x^*,y^*,z^*,\Sigma^*,\Omega_k^*,\Omega_s^*
        \right) =
        \left(
            \tfrac{\sqrt{6}\gamma_m}{2\lambda},\ 
            \sqrt{\tfrac{3\gamma_m(2-\gamma_m)}{2\lambda^2}},\ 
            1 - \tfrac{3\gamma_m}{\lambda^2},\ 
            0,\ 0,\ 0
        \right).
    \end{equation}
    The $(x,y)$ block reproduces the standard CLW eigenvalues, denoted collectively by $\mu_\pm$. The geometric directions give
    \begin{equation}
        \mu_\Sigma = \tfrac{3}{2}(\gamma_m - 2),\qquad
        \mu_k = 3\gamma_m - 2,\qquad
        \mu_s = 3(\gamma_m - 2).
    \end{equation}
    For dust or radiation ($\gamma_m \ge 1$) one has $\mu_k>0$ and $\mu_\Sigma,\mu_s<0$. Shear and spin--torsion are damped, but curvature grows. In the full six-dimensional phase space the scaling solution is a saddle: it is stable in $(x,y,\Sigma,\Omega_s)$ but unstable in the curvature direction \cite{Billyard1998Scaling}.

    \subsection{Summary of fixed points}
    The fixed-point structure is summarized in Table~\ref{tab:extended_fp}. On the $s=+1$ expanding branch and for $\lambda^2<\min(2,3\gamma_m)$, the scalar-dominated point is a stable late-time attractor that drives accelerated expansion. The scaling solution, although attractive in the scalar--fluid subspace, is destabilized by curvature and remains a saddle in the extended Einstein--Cartan phase space. The kinetic points are marginal in the shear and spin directions; homogeneous anisotropy is only genuinely suppressed in the ekpyrotic regime with $w_\phi>1$, not at the exact stiff fixed points themselves.

    \begin{table*}[htbp]
        \centering
        \footnotesize
        \setlength{\tabcolsep}{3pt}
        \resizebox{\textwidth}{!}{
        \begin{tabular}{|c|c|c|c|c|}
            \hline
            \textbf{Point} & \textbf{Coordinates $(x,y,z,\Sigma,\Omega_k,\Omega_s)$} & \textbf{Existence} & \textbf{Eigenvalues $\{\mu\}$} & \textbf{Stability} \\
            \hline
            Kinetic $\pm$ &
            $(\pm 1,0,0,0,0,0)$ & Always &
            $\begin{gathered}
                6,\; 3\mp\sqrt{\tfrac{3}{2}}\,\lambda,\; 3(2-\gamma_m), \\
                0,\; 4,\; 0
            \end{gathered}$ &
            Saddle (marginal in $\Sigma,\Omega_s$) \\
            \hline
            Fluid-dominated &
            $(0,0,1,0,0,0)$ & Always &
            $\begin{gathered}
                \tfrac{3}{2}(\gamma_m - 2),\; \tfrac{3}{2}\gamma_m,\; 3\gamma_m, \\
                \tfrac{3}{2}(\gamma_m - 2),\; 3\gamma_m - 2,\; 3(\gamma_m - 2)
            \end{gathered}$ &
            Saddle \\
            \hline
            Scalar-dominated &
            $\begin{gathered}
                \big(\tfrac{\lambda}{\sqrt{6}},\, \sqrt{1-\tfrac{\lambda^2}{6}}, \\
                0,\, 0,\, 0,\, 0\big)
            \end{gathered}$ & $\lambda^2 < 6$ &
            $\begin{gathered}
                \lambda^2,\; \tfrac{\lambda^2-6}{2},\; \lambda^2-3\gamma_m, \\
                \tfrac{\lambda^2-6}{2},\; \lambda^2-2,\; \lambda^2-6
            \end{gathered}$ &
            Attractor (expanding; $\lambda^2<\min(2,3\gamma_m)$) \\
            \hline
            Scaling &
            $\begin{gathered}
                \big(\tfrac{\sqrt{6}\gamma_m}{2\lambda},\, \sqrt{\tfrac{3\gamma_m(2-\gamma_m)}{2\lambda^2}}, \\
                1-\tfrac{3\gamma_m}{\lambda^2},\, 0,\, 0,\, 0\big)
            \end{gathered}$ & $\lambda^2 > 3\gamma_m$ &
            $\begin{gathered}
                \mu_{\pm},\; \tfrac{3}{2}(\gamma_m - 2), \\
                3\gamma_m - 2,\; 3(\gamma_m - 2)
            \end{gathered}$ &
            Saddle (unstable in $\Omega_k$) \\
            \hline
        \end{tabular}
        }
        \caption{Fixed points and linear stability of the extended dynamical system in $(x,y,z,\Sigma,\Omega_k,\Omega_s)$. The kinetic points are marginal in the shear and spin directions, but the associated ekpyrotic regime ($w_\phi>1$) still suppresses anisotropy along the contracting branch. At the scalar-field-dominated point, the eigenvalue $\lambda^2$ is orthogonal to the Friedmann constraint surface and is not part of the physical stability spectrum on the constrained manifold; the stability classification is based only on the remaining eigenvalues tangent to the constraint surface. Stability is defined with respect to $N=\ln a$; along the contracting branch ($H<0$), $N$ decreases, so a positive eigenvalue $\mu>0$ still corresponds to decay of the associated perturbation. For the scaling point, only the five physical eigenvalues tangent to the constraint surface are listed; the sixth, orthogonal eigenvalue is omitted.}
        \label{tab:extended_fp}
    \end{table*}
    \FloatBarrier

    \subsection{Lyapunov exponent and nonlinear stability}
    \label{sec:lyapunov}
\begin{figure}[t]
        \centering
        \includegraphics[width=0.75\linewidth]{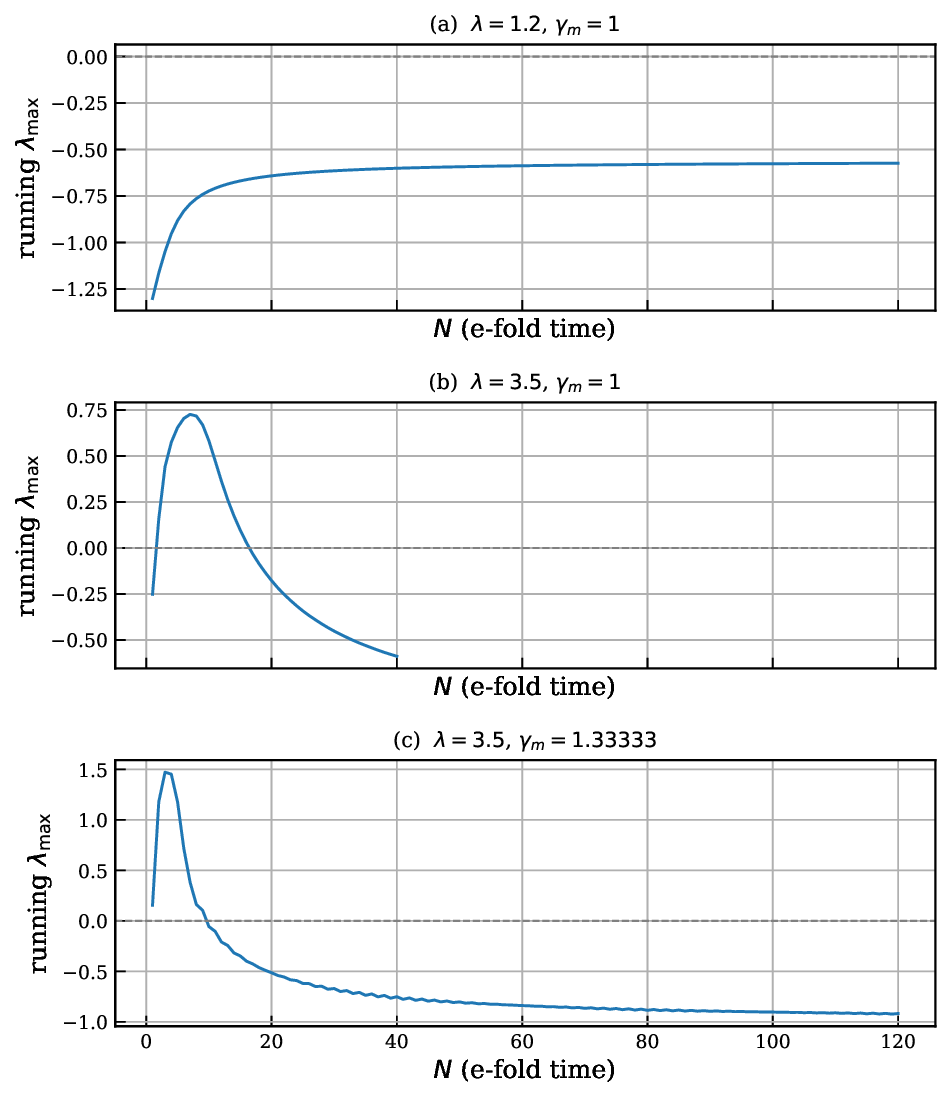}
        \caption{Running maximal Lyapunov exponent $\lambda_{\max}(N)$ for the extended Einstein--Cartan system, computed with the Benettin
            algorithm (Sec.~\ref{sec:lyapunov}). Panels show
            (a) $(\lambda,\gamma_m) = (1.2,1)$,
            (b) $(\lambda,\gamma_m) = (3.5,1)$, and
            (c) $(\lambda,\gamma_m) = (3.5,4/3)$. In all cases $\lambda_{\max}$ tends toward a negative plateau on the expanding branch, indicating no evidence of chaotic mixing in these homogeneous backgrounds for the constant-slope ($\lambda=\mathrm{const}$) slices explored here.}
        \label{fig:lyapunov_panels}
    \end{figure}

    To probe stability beyond the linear regime I compute the maximal Lyapunov exponent $\lambda_{\max}$ using the Benettin algorithm \cite{Benettin1980a,Benettin1980b} in the continuous-time formulation of Ref.~\cite{Cvitanovic2023}. The extended Einstein--Cartan system is six-dimensional, but the Friedmann equation is encoded in the dimensionless density variables, so I eliminate $z$ via the constraint and work with
    \[
        \mathbf{X} = (x,y,\Sigma,\Omega_k,\Omega_s)
    \]
    and independent variable $N = \ln a$. Two nearby trajectories, $u_{\rm ref}(N)$ and $u_{\rm pert}(N)$, are evolved in this five-dimensional phase space; their separation $d(N) = u_{\rm pert}(N) - u_{\rm ref}(N)$ is periodically renormalized to a fixed norm $\delta_0$ after each interval $\Delta N$ in e-fold time. In the runs shown in Fig.~\ref{fig:lyapunov_panels} I take $\delta_0 = 10^{-8}$ and $\Delta N = 1$. The quoted Lyapunov exponents therefore characterize the flow tangent to the Friedmann-constraint surface, rather than off-manifold perturbations in the eliminated $z$-direction. The running estimate of the maximal Lyapunov exponent is
    \begin{equation}
        \lambda_{\max}(N) \approx \frac{1}{N - N_0}\sum_{k=1}^{M}
        \ln\frac{\|d_k\|}{\delta_0},
    \end{equation}
    where $N_0$ is the initial e-fold time and $M$ is the number of renormalization steps up to $N$. The evolution is performed with the explicit eighth-order Runge--Kutta method DOP853 and fixed $(\lambda,\gamma_m)$, corresponding to the exponential part of the potential.

    To keep the spin sector physical I enforce $\Omega_s \ge 0$ by clipping small negative numerical values of $\Omega_s$ to zero. I also monitor the constraint combination
    \begin{equation}
        \mathcal{C}(N) \equiv x^2 + s\,y^2 + \Sigma^2 + \Omega_k - \Omega_s,
    \end{equation}
    which equals $1 - z$ when the Friedmann constraint \eqref{eq:Friedmann} holds. Since $z \ge 0$ in the physical domain, one must have $\mathcal{C}(N) \le 1$. If $\mathcal{C}$ ever exceeds $1 + 10^{-6}$ the integration is flagged as unreliable and terminated. For all runs reported here the trajectories stay within this bound for the full integration interval. Apart from clipping small negative $\Omega_s$ values, I do not project the solution back into the interior of the domain, so the Lyapunov exponents probe the flow of the reduced system with $\Omega_s \ge 0$.

    Representative running exponents for three parameter choices are shown in Fig.~\ref{fig:lyapunov_panels}. For $(\lambda,\gamma_m) = (1.2,1.0)$, the running Lyapunov exponent quickly settles to a negative plateau, $\lambda_{\max} \simeq -0.58$ for $N \gtrsim 40$, and stays close to this value out to $N \approx 120$. For $(\lambda,\gamma_m) = (3.5,1.0)$, there is a brief positive excursion at early times, $\lambda_{\max}(N\!\approx\!10) \sim 0.6$, after which the running exponent relaxes to a similar negative value, $\lambda_{\max} \simeq -0.6$ by $N \approx 40$. This early bump is a finite-time effect; the late-time plateau is independent of the particular norm on phase space (for fixed $N$ as the time variable) and is what I use to characterize nonlinear stability. For $(\lambda,\gamma_m) = (3.5,4/3)$, the convergence is even sharper: the running exponent crosses zero quickly and then decreases toward $\lambda_{\max} \simeq -0.9$ by $N \approx 100$. Varying the renormalization interval $\Delta N$, the integration timestep, and the initial deviation direction within reasonable ranges shifts these plateaus by at most $\mathcal{O}(0.05)$ but never changes their sign. Therefore, the conclusion $\lambda_{\max}<0$ for the expanding, constant-slope slices explored here (integrated forward in $N$) is robust.

    The ekpyrotic contracting regime ($V<0$, $w_\phi \gg 1$) is not directly sampled by the forward-$N$ Lyapunov runs, which mainly probe the expanding branch. Its homogeneous stability is already fixed by the linear analysis: as shown in Secs.~\ref{sec:shear_suppression} and~\ref{sec:eigenvalues}, the shear mode has an effective eigenvalue $\mu_\Sigma^{\rm eff} \simeq \tfrac{3}{2}(w_\phi-1)>0$ in the $N$-time formulation, so the ekpyrotic regime is attractive as $N\to -\infty$ and homogeneous shear decays exponentially along the contracting branch. No additional Lyapunov computation is needed to establish homogeneous ekpyrotic stability. Together with the negative maximal Lyapunov exponents found on the expanding branch, this gives no evidence for chaotic mixing in the homogeneous Einstein--Cartan backgrounds studied here within this truncated phase space and for the parameter ranges probed. This conclusion is restricted to the homogeneous sector; BKL-type chaotic behavior may still arise once spatial inhomogeneities are included.

   \section{Numerical phase-space trajectories and torsion bounce}
   \label{sec:numerics}
   To complement the fixed-point and linear stability analysis in Secs.~\ref{sec:shear_suppression}--\ref{sec:eigenvalues}, I now integrate the extended Einstein--Cartan system directly. I use $N=\ln a$ to study late-time attractors of the autonomous system, and cosmic time $t$ to resolve the nonsingular bounce, where the normalized variables become ill-defined as $H\to 0$.

    Phase-space trajectories and Lyapunov evolution in $N$-time are computed with the DOP853 integrator. The stiff background evolution through the torsion-induced bounce in cosmic time is integrated with an implicit Radau IIA method.

    \subsection{Phase-space trajectories in $(x,y,z)$}
    \label{sec:numerics_phase}
\begin{figure}[htpb!]
        \centering
        \subfloat[\label{fig:phase_xy}]{
            \includegraphics[width=0.78\linewidth]{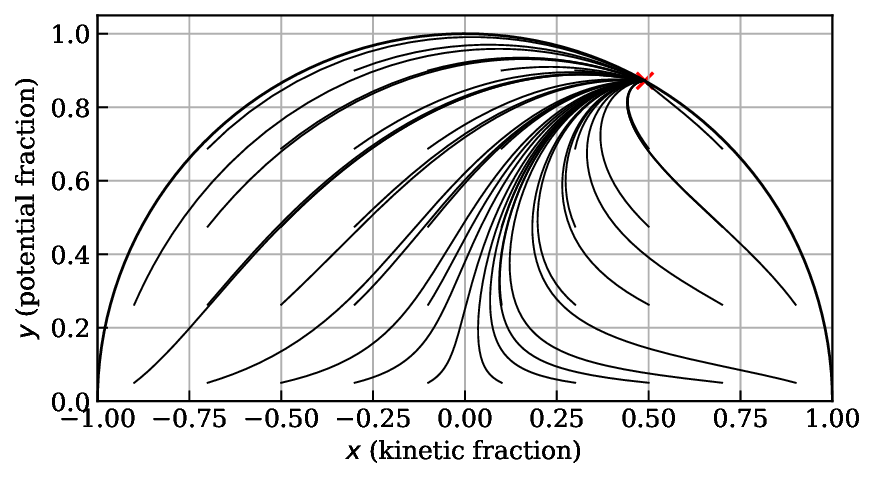}}
        \par\vspace{0.2cm}
        \subfloat[\label{fig:phase3d}]{
            \includegraphics[width=0.78\linewidth]{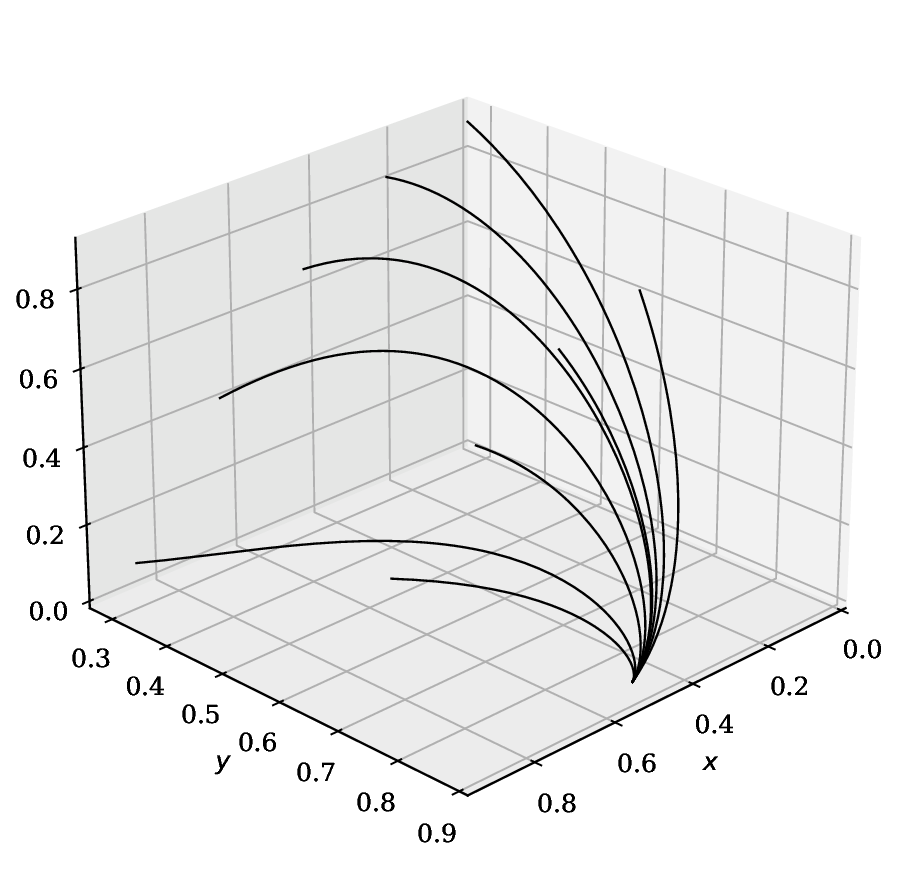}}
        \caption{Phase-space structure of the expanding $s=+1$ branch for $(\lambda,\gamma_m)=(1.20,1.00)$. 
        (a) Projection onto $(x,y)$. Starting from small anisotropy, curvature, and spin--torsion seeds, trajectories inside the allowed domain converge to the scalar-field dominated expansion fixed point $(x_*,y_*)$ identified in the linear stability analysis. The torsion-regulated bounce lies in the $V<0$ branch and is not part of this constant-$\lambda$ slice. 
        (b) Three-dimensional flow in $(x,y,z)$ for the same run, showing the decay of the fluid fraction $z$ as the orbit approaches the scalar attractor.}
        \label{fig:ecem_intro_phase}
    \end{figure}

    To visualize the expanding branch, I fix $(\lambda,\gamma_m)=(1.20,1.0)$ and approximate the potential locally as exponential so that $\lambda$ is constant and $s=+1$. Numerical trajectories in $(x,y,z)$ for this constant-$\lambda$ slice are shown in Fig.~\ref{fig:ecem_intro_phase} and agree with the fixed-point picture in Sec.~\ref{sec:eigenvalues}. These runs probe only the $s=+1$ expanding regime; the dependence of full expand--contract--bounce histories on the softening parameters $(\phi_{\rm b},\Delta)$ is analyzed in Sec.~\ref{sec:bounce-basin}.

    \begin{itemize}
        \item On the expanding branch, orbits with small initial anisotropy, curvature, and spin--torsion spiral toward the scalar-field dominated fixed point found in Sec.~\ref{sec:eigenvalues}, with $z\to 0$ and $(x,y)$ approaching the analytic coordinates, as in Fig.~\ref{fig:ecem_intro_phase}(a).
        \item The three-dimensional trajectories in Fig.~\ref{fig:ecem_intro_phase}(b) show the same flows in $(x,y,z)$, making the decay of the fluid fraction $z$ explicit as the orbit approaches the scalar attractor.
    \end{itemize}

    \FloatBarrier

    \subsection{Bounce in cosmic time}
    \label{sec:bounce_in_time}
    The normalized variables $(x,y,z,\Sigma,\Omega_k,\Omega_s)$ provide a dimensionless phase-space description that is convenient for studying asymptotic behavior on the expanding and contracting branches. Because they divide explicitly by $H$, however, the torsion-regulated bounce cannot be followed directly in this formulation: as $H\to 0$, the normalized variables diverge and the autonomous system hits a coordinate singularity at $H=0$.

\begin{figure}[htpb!]
        \centering
        \subfloat[\label{fig:bounce}]{
            \includegraphics[width=0.85\linewidth]{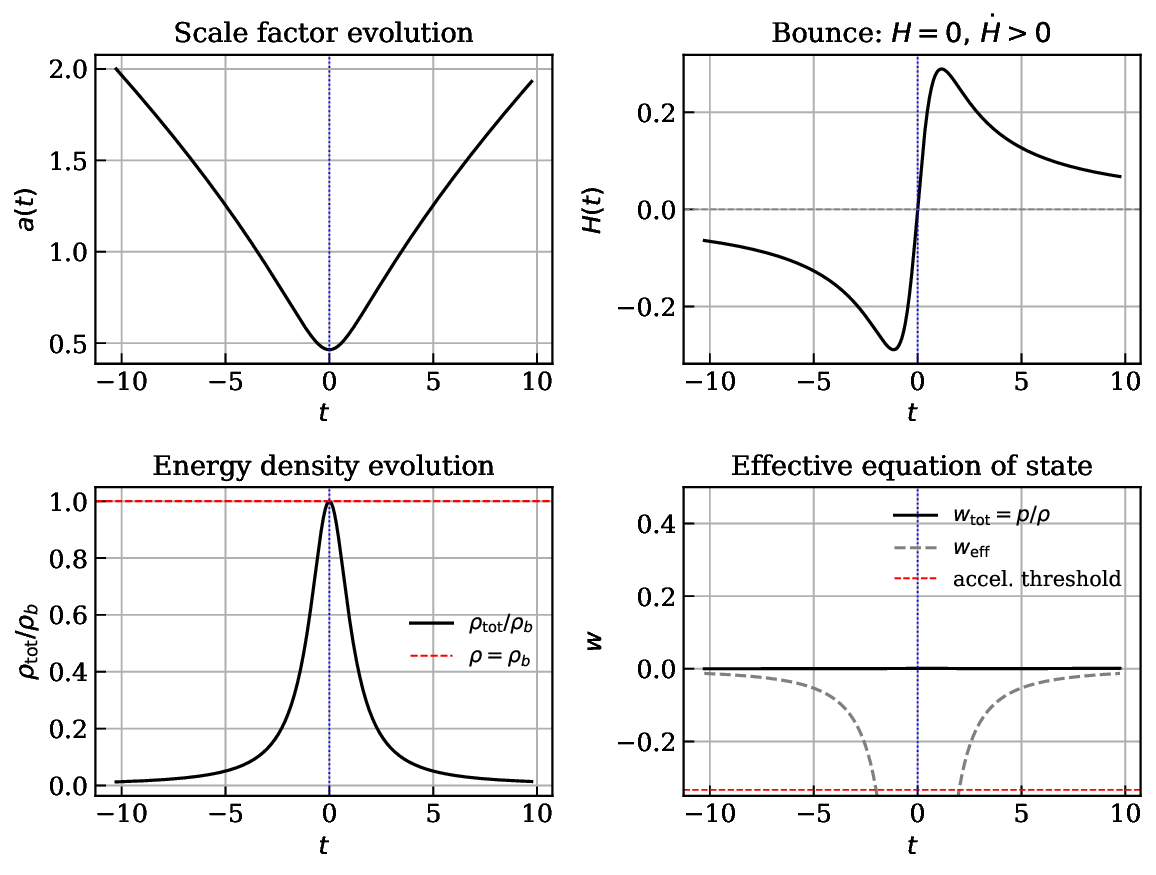}}
        \par\vspace{0.2cm}
        \subfloat[\label{fig:bounce_error}]{
            \includegraphics[width=0.85\linewidth]{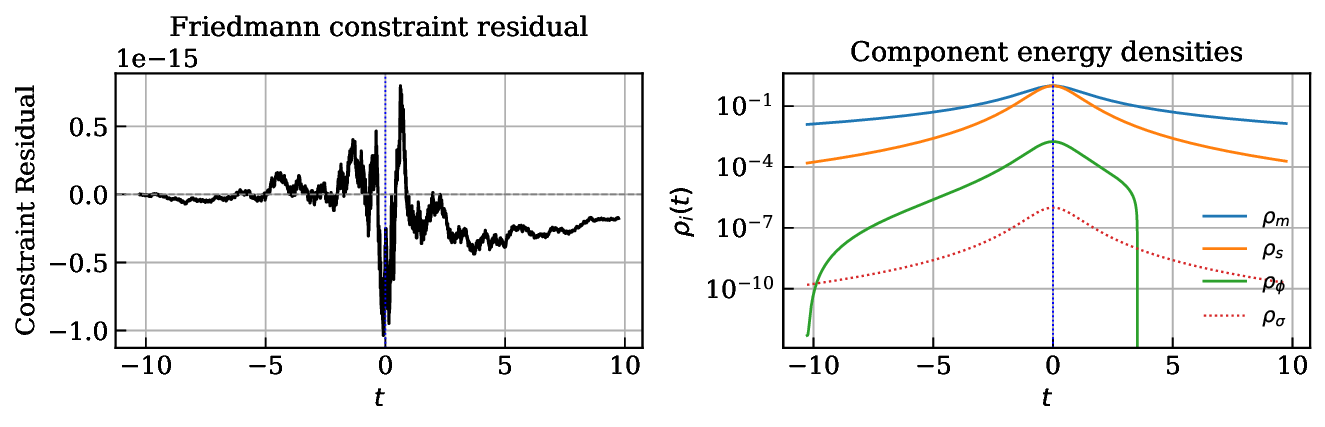}}
        \caption{Representative torsion-regulated Einstein--Cartan bounce and numerical diagnostics for a spatially flat ($k=0$) benchmark run. 
        (a) Cosmic-time evolution of the scale factor, Hubble parameter, normalized total density 
        $\rho_{\rm tot}/\rho_{\rm b}$, and equation-of-state parameters for a contracting solution that undergoes a nonsingular bounce governed by the modified Friedmann equation~\eqref{eq:EC-Friedmann}. The universe contracts from a dust-dominated phase to a finite minimum scale factor $a_{\rm b}$ at $t=0$, where $H$ crosses zero with $\dot H>0$. The total fluid equation of state $w_{\rm tot}=p_{\rm tot}/\rho_{\rm tot}$ approaches dust far from the bounce, while the effective equation of state $w_{\rm eff}$ inferred from $H(t)$ briefly drops below $-1/3$ in a neighborhood of the bounce (where $H\neq 0$), reflecting torsion-driven acceleration rather than exotic matter in the physical sector.
        (b) Left: Friedmann-constraint residual along the same solution, which remains close to zero throughout the evolution. 
        Right: component energy densities on a logarithmic scale, showing that background matter dominates far from the bounce, while the repulsive spin--torsion density $\rho_s$ becomes comparable to $\rho_{\rm tot}$ at the bounce, with scalar and shear contributions remaining subdominant. Because this panel is logarithmic, any non-positive $\rho_\phi$ values on the late post-bounce tail (from the softened potential's negative branch) are not displayed.}
        \label{fig:bounce_results}
    \end{figure}

    To track the bounce I instead work in cosmic time $t$ with unnormalized background variables $(a,H,\phi,\dot{\phi})$ and integrate the Einstein--Cartan Friedmann system~\eqref{eq:EC-Friedmann}. The total energy density
    \[
        \rho(a,\phi,\dot\phi) = \rho_\phi + \rho_m + \rho_\sigma
    \]
    includes a canonical scalar with the softening exponential potential, a perfect fluid with barotropic index $\gamma_m$, and a small shear contribution $\rho_\sigma \propto a^{-6}$. Torsion enters through the stiff $a^{-6}$ correction $\rho_s$ in Eq.~\eqref{eq:EC-Friedmann}. For reference I define a critical density $\rho_{\rm b}$ as the maximum total density reached at the bounce,
    \[
        \rho_{\rm b} \equiv \rho_{\rm tot}(t_{\rm b}),
    \]
    which equals $\rho_s(t_{\rm b})$ for $k=0$ by Eq.~\eqref{eq:EC-Friedmann}. In this $(a,H,\phi,\dot{\phi})$ description the EC equations are regular at $H=0$, and the bounce is characterized simply by
    \begin{equation}
        H(t_{\rm b}) = 0,\qquad \dot H(t_{\rm b}) > 0,
    \end{equation}
    matching the usual conditions for a nonsingular bounce in GR-like cosmologies \cite{Ijjas2018Bouncing}. The normalized dynamical-system variables are used to describe how trajectories approach and exit the high-density regime. Each turning point $H=0$ is then resolved in cosmic time. In practice, I use normalized variables for branch-wise stability and integrate the unnormalized cosmic-time system through the $H=0$ crossings, both in the $k=0$ bounce benchmark of Fig.~\ref{fig:bounce_results} and in the $k=+1$ cyclic trajectory of Fig.~\ref{fig:ecem_cycle}.

    For the trajectory in Fig.~\ref{fig:bounce_results}, the initial data are chosen on the contracting branch ($H<0$) with small shear and spatially flat curvature ($k=0$). This benchmark run keeps $k=0$ throughout. Because $\Omega_k=0$ is an invariant manifold of the normalized system ($\Omega_k' = 2q\,\Omega_k$), this flat benchmark cannot recollapse; expansion-to-contraction turnaround is shown separately in the closed ($k=+1$) trajectory of Fig.~\ref{fig:ecem_cycle}. The solution contracts until the total density approaches $\rho_b$ at a finite scale factor $a_{\rm b}$, at which point $H(t)$ changes sign while $a(t)$ stays nonzero, realizing a smooth torsion-regularized bounce.

    Figure~\ref{fig:bounce_results} shows a representative evolution. Panel~(a) summarizes the background across the bounce: the upper panels show $a(t)$ and $H(t)$, and the lower panels show the total density $\rho(t)$ relative to $\rho_{\rm b}$ and the equation-of-state parameters $w_{\rm tot}$ and $w_{\rm eff}$. Here $w_{\rm tot} \equiv p_{\rm tot}/\rho_{\rm tot}$, and away from the exact bounce point where $H=0$ I define
    \[
        w_{\rm eff} \equiv -1 - \frac{2}{3}\,\frac{\dot H}{H^2},
    \]
    the GR-like effective equation-of-state parameter one would infer from $H^2 = \rho_{\rm eff}/3$. The scale factor reaches a nonzero minimum while $H$ crosses zero smoothly with $\dot H>0$, giving a nonsingular Einstein--Cartan bounce. Panel~(b) shows the Friedmann-constraint residual over the same interval together with the component densities: the residual stays small throughout, and the densities illustrate how matter dominates far from the bounce while the spin--torsion term becomes comparable to the total density at $t_{\rm b}$.

    The existence of torsion-regularized bounces is not limited to this single example. The dependence on the softening parameters $(\phi_{\rm b},\Delta)$ is studied in Sec.~\ref{sec:bounce-basin} and summarized by the bounce-basin plot in Fig.~\ref{fig:bounce_basin}.

    \FloatBarrier

    \section{Implications and discussion}
    \label{sec:implications}
    The phase-space analysis and representative integrations show that, for suitable choices of the spin parameter $\alpha$ and of the softening sector, the homogeneous Einstein--Cartan system with a canonical scalar field admits a \emph{class} of nonsingular background histories in which expansion, fluid domination, ekpyrotic contraction, and a torsion-regulated bounce occur in sequence. Figure~\ref{fig:bounce_results} presents the spatially flat ($k=0$) benchmark that isolates the bounce mechanism. Figure~\ref{fig:ecem_cycle} gives a positively curved ($k=+1$) trajectory with repeated turnaround and bounce crossings. In this restricted sense the construction is a direct extension of familiar quintessence dynamics \cite{Chevallier2001,Linder2003}: a single scalar field drives both late-time acceleration and ekpyrotic smoothing at the background level, while the spin--torsion sector supplies the repulsive $\rho_s\propto a^{-6}$ contribution that regulates the high-density regime, instead of introducing separate ingredients for each phase. Throughout I model the spin--torsion contribution as an $a^{-6}$ component parameterized by $\alpha$ and I do \emph{not} claim that this phenomenological setup captures the full microscopic content of Einstein--Cartan theory or describes generic initial data.

    The setup is intentionally conservative. Any background-level departure from $\Lambda$CDM changes both distances and growth, so late-time parameter inferences are in principle sensitive to these dynamics \cite{Planck2018Parameters}. The ECEM background does not contain an explicit early-dark-energy component, and I do not attempt parameter estimation. In any scenario with repeated expand--contract--bounce episodes, integrated quantities (comoving distances, accumulated lensing, etc.) can carry memory of earlier phases, so fitting $\Lambda$CDM-based templates to such a background could bias inferred parameters \cite{Abdalla2022Tensions}. At this stage, ECEM should be viewed as a model-building framework at the homogeneous level; whether it alleviates or worsens specific tensions can only be decided once perturbations are included and the model is compared directly with data.

    \subsection{Physical scales and validity limits}
    \label{sec:phys_scales}
    A basic consistency requirement is that the torsion-induced bounce occurs parametrically below the Planck density, so that higher-curvature and higher-torsion corrections remain subdominant. Restoring units for this estimate, the reduced Planck mass is $M_{\rm Pl}\simeq 2.4\times 10^{18}\,\mathrm{GeV}$, with Planck density $\rho_{\rm Pl}\sim M_{\rm Pl}^4$. I choose a fiducial bounce scale $E_{\rm b}\sim 10^{16}\,\mathrm{GeV}$, representative of a grand-unified threshold, and define
    \begin{equation}
        \rho_{\rm b} \equiv \rho_{\rm tot}(t_{\rm b}) \simeq E_{\rm b}^4 \sim 10^{64}\,\mathrm{GeV}^4
    \end{equation}
    as the physical (non-spin) energy density at the bounce time $t_{\rm b}$ where $H(t_{\rm b})=0$; for $k=0$ this equals $\rho_s(t_{\rm b})$ by Eq.~\eqref{eq:EC-Friedmann}. The ratio to the Planck density is
    \begin{equation}
        \frac{\rho_{\rm b}}{\rho_{\rm Pl}}
        \sim \left(\frac{E_{\rm b}}{M_{\rm Pl}}\right)^4
        \sim 10^{-10},
    \end{equation}
    so the bounce lies comfortably below $\rho_{\rm Pl}$ and in the same general regime usually assumed in semi-classical bounce scenarios such as loop quantum cosmology \cite{AshtekarSingh2011}. In the $8\pi G=1$ units used elsewhere this corresponds to $\rho_{\rm b}\sim 10^{-10}$ in Planck units, and with $a(t_{\rm b})=1$ the parameter $\alpha$ in $\rho_s\equiv\alpha a^{-6}$ is fixed to $\alpha\simeq\rho_{\rm b}$.

    In a microscopic Einstein--Cartan treatment, the spin--torsion contribution for a fermion fluid can be written schematically as
    \begin{equation}
        \rho_s \sim \xi\,\frac{n^2}{M_{\rm Pl}^2},
    \end{equation}
    where $n$ is the fermion number density and $\xi=\mathcal{O}(1)$ depends on spin representation and conventions \cite{Weyssenhoff1947,Hehl1976Torsion,Shapiro2002Torsion,Poplawski2010Torsion}. Imposing $\rho_s\simeq\rho_{\rm b}$ at the bounce gives
    \begin{equation}
        n_{\rm b} \sim \sqrt{\rho_{\rm b}}\,M_{\rm Pl} \sim E_{\rm b}^2 M_{\rm Pl} \sim 10^{50}\,\mathrm{GeV}^3
    \end{equation}
    for $E_{\rm b}\sim 10^{16}\,\mathrm{GeV}$. A relativistic fermion bath in equilibrium at $T\sim E_{\rm b}$ has $n_{\rm th}\sim g_* T^3$, giving $n_{\rm th}\sim 10^{50}\,\mathrm{GeV}^3$ for $g_*\sim 10^2$. The required spin density is of the same order as the thermal expectation, consistent with a mildly overdense or degenerate fermion fluid rather than a trans-Planckian state.

    These back-of-the-envelope estimates indicate that the fiducial bounce scale $\rho_{\rm b}\sim 10^{-10}\rho_{\rm Pl}$ is at least \emph{not obviously incompatible} with reasonable fermion number densities in simple Weyssenhoff-type fluids. In what follows I treat $E_{\rm b}$ and the associated $\alpha\simeq\rho_{\rm b}$ as parametrizing the regime of validity of the effective Einstein--Cartan description and do not attempt to model loop-induced higher-curvature or higher-torsion operators, which could become important closer to $\rho_{\rm Pl}$.

    \subsection{Anisotropy, curvature, and parameter constraints}
    Dynamically, the background in this model spends most of its time close to FLRW. The main issue in contraction is the fate of shear. Purely kinetic-dominated solutions with $w_\phi = 1$ behave as stiff fluids and leave the shear eigenvalue $\mu_\Sigma = 0$, so $\Sigma$ is only marginally stable. By contrast, as shown in Sec.~\ref{sec:shear_suppression}, a steep negative potential with $w_\phi \gg 1$ makes the ekpyrotic regime a strong homogeneous attractor \cite{Erickson2004Ekpyrotic,Garfinkle2008Ekpyrotic}:
    \[
        \Sigma \propto e^{(q-2)N} \to 0
    \]
    along the contracting branch. For initial data with small but nonzero shear, the torsion-driven bounce is approached in a nearly isotropic homogeneous state.

    Both shear and spin--torsion contributions behave as stiff components,
    \[
        \rho_\sigma \propto a^{-6},\qquad \rho_s \propto a^{-6},
        \qquad\Rightarrow\qquad
        \frac{\rho_\sigma}{\rho_s} = \text{const}.
    \]
    In the ECEM background, the ekpyrotic phase does \emph{not} generate a hierarchy between shear and spin: any hierarchy must be present in the initial data and is then preserved as the universe contracts. Consequently, the bounce is only safe from shear domination if the initial stiff sector is already spin-dominated, $\rho_\sigma \ll \rho_s$; in that case the hierarchy is preserved throughout contraction. If instead the initial data satisfy $\rho_\sigma \gtrsim \rho_s$, the high-density regime would be shear-dominated and the simple homogeneous ECEM truncation used here would no longer be self-consistent near the would-be bounce. What ekpyrosis does accomplish is to suppress both shear and spin relative to the scalar field. During ekpyrotic contraction with $w_\phi \simeq w_{\rm ek}>1$ one has
    \[
        \rho_\phi \propto a^{-3(1+w_{\rm ek})},
    \]
    so that
    \[
        \frac{\rho_\sigma}{\rho_\phi}(N_{\rm soft})
        \;\sim\;
        \frac{\rho_\sigma}{\rho_\phi}\big|_{\rm in}\,
        \exp\!\Big[-3\big(w_{\rm ek}-1\big)N_{\rm ek}\Big],
    \]
    where "in" denotes the start of the ekpyrotic phase, $N_{\rm ek} = N_{\rm in} - N_{\rm soft} > 0$ is the absolute number of e-folds of ekpyrotic contraction, and $N_{\rm soft}$ marks the onset of the softening regime. Demanding
    $\rho_\sigma/\rho_\phi(N_{\rm soft})\lesssim\varepsilon$ gives a shear-suppression bound
    \begin{equation}
        N_{\rm ek} \;\gtrsim\; \frac{1}{3\big(w_{\rm ek}-1\big)}\,
        \ln\!\left[\frac{\big(\rho_\sigma/\rho_\phi\big)_{\rm in}}{\varepsilon}\right].
    \end{equation}
    In the parameter choices used here, the bounce occurs when the spin sector becomes comparable to the scalar, $\rho_s \sim \rho_\phi$. Therefore, if the initial data satisfy $\rho_\sigma \ll \rho_s$ in the stiff sector and the ekpyrotic phase lasts long enough that $\rho_\sigma \ll \rho_\phi$ by the onset of softening, then shear remains subdominant to both dominant components at the bounce.

    The trajectories in Fig.~\ref{fig:ecem_intro_phase} and the bounce solution in Fig.~\ref{fig:bounce_results} are consistent with this picture: the expanding branch approaches the scalar attractor with $z\to 0$, while the bounce occurs in a regime where the spin sector is comparable to the total density, the shear fraction remains small throughout, and $\rho_\sigma/\rho_s$ stays approximately constant.

    Spatial curvature behaves similarly. Near the scaling saddle, curvature is a growing mode with $\mu_k = 3\gamma_m - 2 > 0$ for $\gamma_m \ge 1$ \cite{Billyard1998Scaling}, but once the ekpyrotic field dominates it is strongly suppressed. For approximately constant equation of state $w$, the curvature variable obeys
    \begin{equation}
        \Omega_k' = 2q\,\Omega_k = (1+3w)\,\Omega_k.
    \end{equation}
    Thus $\Omega_k=0$ is an invariant manifold of the normalized system: a spatially flat ($k=0$) trajectory cannot dynamically develop curvature or recollapse, so the expansion-to-contraction turnaround requires the closed branch ($k=+1$).
    During a standard fluid-dominated \emph{expansion} with $w_{\rm sc}\in[0,1/3]$ and duration $N_{\rm sc}$,
    \begin{equation}
        |\Omega_k|_{\rm after\;sc} \;\sim\; |\Omega_k|_{\rm in}\,
        \exp\!\big[(1+3w_{\rm sc})N_{\rm sc}\big],
    \end{equation}
    with $(1+3w_{\rm sc})\in[1,2]$. During ekpyrotic \emph{contraction} with $w_{\rm ek}>1$ and duration $N_{\rm ek}$,
    \begin{equation}
        |\Omega_k|_{\rm after\;ek} \;\sim\; |\Omega_k|_{\rm after\;sc}\,
        \exp\!\big[-(1+3w_{\rm ek})N_{\rm ek}\big],
    \end{equation}
    because $N$ decreases by $N_{\rm ek}$ as $a$ shrinks. Combining these and demanding $|\Omega_k|_{\rm after\;ek}\lesssim\varepsilon_k$ gives the curvature-basin inequality
    \begin{equation}
        (1+3w_{\rm ek})N_{\rm ek}
        \;\gtrsim\;
        (1+3w_{\rm sc})N_{\rm sc}
        + \ln\!\left[\frac{|\Omega_k|_{\rm in}}{\varepsilon_k}\right].
    \end{equation}
    For representative values $N_{\rm sc}\sim 60$, $w_{\rm sc}\lesssim 1/3$, and a steep ekpyrotic phase with $w_{\rm ek}\gg 1$, the prefactor $(1+3w_{\rm ek})^{-1}$ is small, so only $N_{\rm ek}\sim\mathcal{O}(1)$--$\mathcal{O}({\rm few})$ e-folds are needed to push curvature back into the observational basin even if it grows substantially during a long standard expansion epoch.

    Over a single expand--contract--bounce episode, anisotropy and curvature can grow transiently but are driven small again before the bounce. This occurs whenever the ekpyrotic phase satisfies the shear- and curvature-bounds above and the initial spin-shear hierarchy is imposed. Within this class of solutions the isotropic manifold is a local attractor during ekpyrotic contraction and remains dynamically stable through the torsion-induced bounce at the homogeneous level. This kind of homogeneous attractor behavior, together with a pre-imposed spin-dominated stiff sector, is a necessary but not sufficient condition for any nonsingular completion of ekpyrosis: without it, one would generically approach the high-density regime in a strongly anisotropic state where the bounce dynamics are uncontrolled. A full inhomogeneous (BKL-type) analysis with Weyssenhoff torsion is left for future work and may well tighten these requirements.

    \subsection{Basin of viability for the softening--spin bounce}
    \label{sec:bounce-basin}
    Given initial data in which the spin--torsion sector is subdominant, the key question is whether the stiff term $\rho_s$ can grow sufficiently relative to the scalar to trigger a torsion-dominated bounce before the universe reaches a singularity. The relevant quantity is the ratio
    \[
        R \;\equiv\; \frac{\rho_s}{\rho_\phi}.
    \]
    Using $\rho_s \propto a^{-6}$ and $\rho_\phi \propto a^{-3(1+w_\phi)}$, 
    \begin{equation}
        \frac{R_2}{R_1}
        \;=\;
        \left(\frac{a_2}{a_1}\right)^{-6 + 3(1+w_\phi)}
        \;=\;
        \exp\!\big[3(1-w_\phi)\,N_{12}\big],
        \label{eq:R_scaling}
    \end{equation}
    where $N_{12} \equiv \ln(a_1/a_2)>0$ is the number of e-folds of contraction between the two epochs.

\begin{figure}[htpb!]
        \centering
        \includegraphics[width=0.75\linewidth]{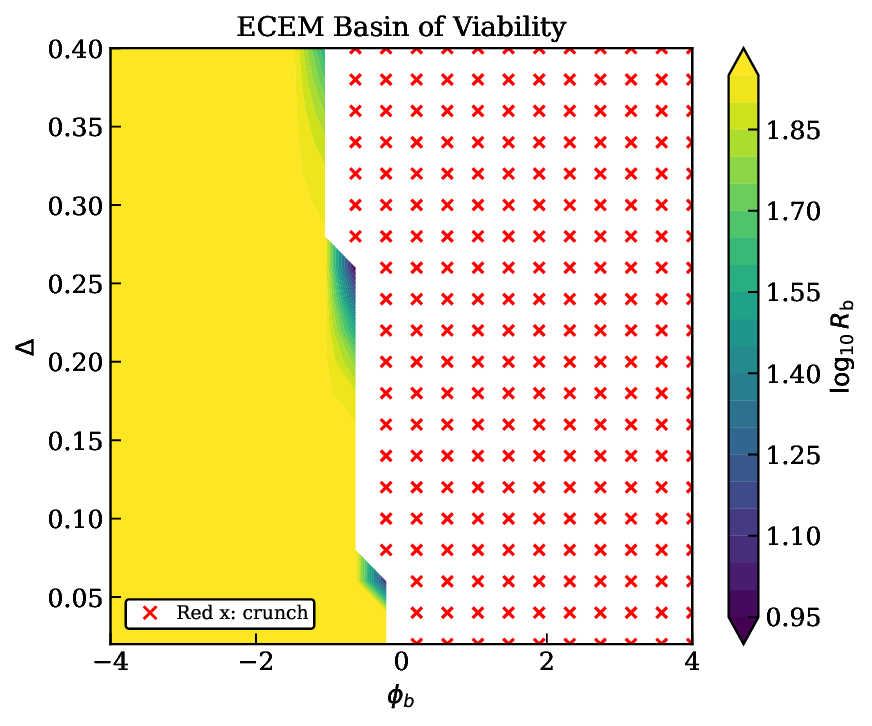}
        \caption{Basin of viability for the softening--spin bounce in the $(\phi_{\rm b},\Delta)$ plane for fixed $(\alpha,V_{\rm soft})$ and fixed initial data. The color scale shows the spin-to-scalar ratio $R_{\rm b} \equiv \rho_s/\rho_\phi$ evaluated at the bounce for trajectories that undergo a torsion-induced reversal of contraction ($H=0,\dot H>0)$; trajectories that instead reach the crunch condition are marked with red $\times$ symbols. To isolate the scalar--spin competition, these basin runs use $\rho_{m0}=3\times10^{-4}$, $\alpha=10^{-4}$, and $a_0=3$, distinct from the high-density, dust-dominated single-bounce example discussed earlier. On the $20\times20$ grid shown here (with the same run settings), the scan yields 176 bouncing trajectories, 224 crunch trajectories, and 0 undecided trajectories. The solid curve marks the numerically determined boundary between bouncing and non-bouncing trajectories and lies close to the contour $R_{\rm b}\simeq 1$ suggested by \eqref{eq:bounce_inequality}: parameter combinations on and above the curve satisfy the softening condition and realize a torsion-dominated bounce within the Einstein--Cartan regime, while points below the curve do not achieve a softening--spin bounce and typically evolve toward a singular crunch.}
        \label{fig:bounce_basin}
    \end{figure}

    Consider three stages:\\
    (i) the onset of ekpyrotic contraction at $(\rho_s/\rho_\phi)_{\rm in} = R_{\rm in}$,\\
    (ii) the beginning of the softening regime after $N_{\rm ek}$ e-folds of ekpyrotic contraction, and\\
    (iii) the bounce. Approximating the ekpyrotic phase by a constant $w_\phi \simeq w_{\rm ek}>1$ and the softening regime by an average equation of state $\bar{w}_{\rm soft}(\phi_b,\Delta,\alpha)$ over a duration $N_{\rm soft}(\phi_b,\Delta,\alpha)$, Eq.~\eqref{eq:R_scaling} gives
    \begin{align}
        R_{\rm soft}
        &\;\simeq\;
        R_{\rm in}\,
        \exp\!\Big[3\big(1-w_{\rm ek}\big)N_{\rm ek}\Big],
        \\
        R_{\rm b}
        &\;\simeq\;
        R_{\rm soft}\,
        \exp\!\Big[3\big(1-\bar{w}_{\rm soft}\big)N_{\rm soft}\Big],
    \end{align}
    so that
    \begin{equation}
        \ln R_{\rm b}
        \;\simeq\;
        \ln R_{\rm in}
        + 3\big(1-w_{\rm ek}\big)N_{\rm ek}
        + 3\big(1-\bar{w}_{\rm soft}\big)N_{\rm soft}.
        \label{eq:Rb_log}
    \end{equation}
    In the ECEM system the torsion-induced bounce occurs when the spin sector grows large enough to offset the scalar contribution in the Friedmann constraint; since matter and shear are subdominant in this regime, the bounce condition reduces to $\rho_s \approx \rho_\phi$, i.e.\ $R_{\rm b} \approx 1$. For fixed initial data $R_{\rm in}$ and ekpyrotic duration $N_{\rm ek}$, this defines a simple "bounce versus crunch" inequality for the softening sector,
    \begin{equation}
        \big(1-\bar{w}_{\rm soft}\big)\,N_{\rm soft}(\phi_b,\Delta,\alpha)
        \;\gtrsim\;
        \big(w_{\rm ek}-1\big)N_{\rm ek}
        + \frac{1}{3}\,\ln\!\left[\frac{1}{R_{\rm in}}\right].
        \label{eq:bounce_inequality}
    \end{equation}
    If \eqref{eq:bounce_inequality} holds, the spin term becomes comparable to the scalar term, $|\rho_s|\sim\rho_\phi$, and offsets the scalar contribution in the Friedmann constraint after the system leaves the pure ekpyrotic attractor, leading to a nonsingular bounce. If it is violated, the scalar remains dominant all the way to $a\to 0$ and the solution collapses to a curvature singularity despite the presence of torsion. In numerical experiments this failure typically occurs in two limiting ways: if the softening transition happens too far above the torsion scale, the spin sector never catches up and the model simply crunches; if it happens too far below, the system has already exited the ekpyrotic attractor and evolves toward a scalar-dominated singularity. The finite "bounce basin" shown in Fig.~\ref{fig:bounce_basin} makes this overlap requirement explicit within the truncated homogeneous model: only parameter choices for which the softening window overlaps the density range where the EC spin term becomes dynamically important lead to a torsion-dominated bounce.

    For the class of profiles used here, the softening regime is characterized by a trajectory-dependent average equation of state $\bar{w}_{\rm soft}<1$ and a duration $N_{\rm soft}$, both determined from the numerical evolution in the reduced scalar--spin sector. Equation~\eqref{eq:bounce_inequality} then picks out a smooth threshold curve in the $(\phi_b,\Delta)$ plane (for fixed $\alpha$) that separates bouncing from non-bouncing solutions. Figure~\ref{fig:bounce_basin} illustrates this by showing contours of $R_{\rm b}$ in the $(\phi_b,\Delta)$ plane: the region above the threshold satisfies \eqref{eq:bounce_inequality} and yields a torsion-induced bounce, while the region below it does not satisfy the softening condition and typically evolves toward a singular crunch (in the absence of additional effects).

    \FloatBarrier

    Thus, once the shear-suppression and curvature-basin bounds are imposed, the existence of a nonsingular EC bounce in the ECEM background is further restricted to a tuned but finite "basin of viability" in the softening parameter space for the specific class of potentials considered here, defined by the condition that the integrated softening,
    \[
        \mathcal{I}_{\rm soft}(\phi_b,\Delta,\alpha)
        \;\equiv\;
        \int_{\rm soft} \big(1-w_\phi\big)\,dN,
    \]
    satisfies
    \[
        \mathcal{I}_{\rm soft}(\phi_b,\Delta,\alpha)
        \;\gtrsim\;
        \mathcal{I}_{\rm crit}
        \;\equiv\;
        (w_{\rm ek}-1)N_{\rm ek}
        + \frac{1}{3}\,\ln\!\left[\frac{1}{R_{\rm in}}\right],
    \]
    so that the softening phase can overcome the prior ekpyrotic suppression of the spin sector. Microscopic ECSK realizations that enforce a specific algebraic relation $\rho_s(\rho_m)$ (for example Weyssenhoff dust or radiation with $\rho_s \propto \rho_m^2$) correspond to lower-dimensional invariant submanifolds of this extended phase space; some such choices may shrink or even eliminate the bounce basin shown in Fig.~\ref{fig:bounce_basin}, and a systematic survey of these submanifolds is left for future work.

    \subsection{Entropy and early structure}
    Entropy production across cycles remains the main conceptual obstacle for any cyclic cosmology. The classic Tolman arguments show that for ordinary matter and radiation, entropy generically increases from cycle to cycle. Therefore, an exactly periodic, thermodynamically eternal sequence of cycles is incompatible with standard thermodynamics unless some entropy-dilution or reset mechanism operates~\cite{Tolman1931Periodic,Tolman1931Entropy}. At the homogeneous level considered here, the Einstein--Cartan bounce is nonsingular and conservative: it modifies the high-density Friedmann equations via the spin--torsion term but does not introduce explicit dissipation or entropy sinks. This removes curvature singularities but does not by itself regulate the total entropy budget. Realistic evolution involves black-hole formation and evaporation, particle production, and other irreversible channels. Some of these channels have been proposed as possible ways to dilute or reset entropy, but none are modeled here. The ECEM background solutions shown here should be viewed as homogeneous templates (including both a $k=0$ single-bounce benchmark and a $k=+1$ cyclic benchmark), not as a complete thermodynamic model of an eternal cyclic universe. If one attempts to embed this type of background in a truly cyclic cosmology, a viable construction would require a microphysical scenario that tracks entropy flow cycle by cycle.

    Similarly, claims of unusually early massive galaxies or rapid structure formation motivate nonstandard expansion histories, but they remain under debate due to selection effects and modeling systematics. The work here is restricted to the homogeneous background and does not evolve perturbations. Questions about the transfer of scalar and tensor modes through the torsion-driven bounce, the resulting power spectra, or any impact on very high-redshift structure and large-scale anomalies are deferred to future work.

    \subsection{Physical interpretation of dynamical regimes}
    Each fixed point or asymptotic regime has the interpretation already summarized in Secs.~\ref{sec:phase-space-structure} and~\ref{sec:numerics}: a scalar-dominated phase with $w_\phi = \lambda^2/3 - 1$ (accelerated for $\lambda^2<2$), fluid-dominated saddles corresponding to the usual matter and radiation eras, and an ekpyrotic contracting regime that suppresses shear anisotropy (Sec.~\ref{sec:shear_suppression}). The softening phase with $w_\phi<1$ is the window in which $\rho_s\propto a^{-6}$ overtakes $\rho_\phi$, allowing $H\to0$ and $\dot H>0$ at a finite $a_{\rm b}$ and producing the nonsingular bounce. This appears as a single event in the $k=0$ benchmark and as a repeated event in the $k=+1$ cyclic trajectory.

    In this description the relation between the softening scale $V_{\rm soft}$ and the spin--torsion parameter $\alpha$ is imposed at the background level rather than derived from microphysics. The potential is chosen so that the transition to the softened plateau occurs near the density at which the EC spin term would otherwise enforce a bounce. At the level of the present model, that alignment should be regarded as a tuning, not as a prediction of a specified ultraviolet completion.

    \section{Observational connections}
    The ECEM background suggests qualitative observational handles at the level of this homogeneous model, even though I do not evolve perturbations, perform data fitting, or derive sharp constraints, and the illustrative histories in Figs.~\ref{fig:ecem_cycle}--\ref{fig:bounce_results} are used only as benchmarks.

    First, because curvature is a growing mode at the scaling saddle, the normalized curvature parameter $\Omega_k(a)$ is generically time-dependent once the background departs from $\Lambda$CDM. Analyses that assume constant curvature but fit data from different redshift ranges can in principle misinterpret an evolving $\Omega_k(a)$. Baryon Acoustic Oscillation and CMB lensing analyses that allow mild time variation in effective curvature \cite{Handley2021Curvature} could be used to test whether a slowly evolving $\Omega_k(a)$ provides a better fit than strictly flat $\Lambda$CDM, or whether current data instead strongly prefer near-constancy of $\Omega_k$ over the relevant redshift window.

    Second, while the baseline ECEM evolution drives large-scale shear to very small values before the bounce, small residual anisotropies and curvature patterns could still imprint the CMB low multipoles or galaxy spin statistics. In the solutions studied here, any surviving anisotropy at late times would, if the background analysis is representative, live on very large, super-horizon scales. Future surveys (e.g.\ JWST, LSST), combined with careful control of selection and pipeline systematics \cite{Shamir2022Spin,Shamir2024JWST}, may provide indirect probes of how close the ECEM background can remain to exact isotropy.

    Third, in many scenarios with a stiff or ekpyrotic phase, the background dynamics can enhance high-frequency stochastic gravitational waves and tilt the tensor spectrum blue \cite{Giovannini1999StiffGW}. Whether an analogous signal arises in the ECEM setup, and with what amplitude, depends on the detailed evolution of tensor perturbations through the torsion-driven bounce and requires a dedicated tensor-perturbation analysis beyond the homogeneous background treated here.

    \section{Conclusions}
    This work constructs an example of a homogeneous Einstein--Cartan background history with a canonical scalar field and small shear, for a restricted class of tuned softening potentials and initial conditions. In this construction, the spin--torsion contribution regulates high densities and connects scalar-dominated expansion, a fluid era, ekpyrotic contraction, and a torsion-induced bounce in a kinematically nonsingular model. A single scalar field with an appropriately chosen softening potential both drives expansion and implements ekpyrotic smoothing, while the spin--torsion sector supplies the repulsive $\rho_s\propto a^{-6}$ component that triggers the bounce. The softening potential resolves the tension between ekpyrotic contraction, where $\rho_\phi$ would otherwise outpace $\rho_s$, and the requirement that the spin--torsion density eventually dominate. In the present construction this is achieved by tuning $(\phi_{\rm b},\Delta,V_{\rm soft})$ rather than deriving these parameters from a microscopic spinor model. In this restricted sense the ECEM construction now demonstrates both: (i) a directly integrated spatially flat ($k=0$) torsion bounce benchmark (Fig.~\ref{fig:bounce_results}), and (ii) a directly integrated closed ($k=+1$) homogeneous trajectory with repeated turnaround and bounce crossings (Fig.~\ref{fig:ecem_cycle}). I do not claim a complete cyclic cosmology or a general resolution of the Big Bang singularity. Whether the universe can execute a thermodynamically eternal sequence of such expand--contract--bounce episodes is a separate question, constrained by the Tolman entropy problem~\cite{Tolman1931Periodic} and modern entropy analyses in cyclic scenarios~\cite{Ijjas2022Entropy}, and lies outside the scope of the present description.

    For the range of initial conditions explored here (with shear and curvature within observational bounds), the trajectories explored numerically remain regular and show no indication of chaotic mixing at the homogeneous level, and the bounce-basin analysis of Sec.~\ref{sec:bounce-basin} shows that these trajectories occupy a tuned but finite region of softening-parameter space rather than an isolated point. The overall bounce scale and initial spin density remain free parameters of the effective description and could in principle be constrained once a concrete perturbation framework and data analysis are specified.

    \subsection{Future work and outlook}
    The next step is to extend the ECEM framework beyond the homogeneous background and assess whether it can serve as a falsifiable background-level alternative to $\Lambda$CDM. This requires evolving scalar and tensor perturbations through the Einstein--Cartan bounce to obtain the scalar power spectrum $P_\zeta(k)$, the tensor tilt $n_T$, and the tensor-to-scalar ratio $r$, and comparing these with CMB and large-scale-structure data. It will also be important to relax strict FLRW symmetry by introducing controlled anisotropy (e.g.\ Bianchi sectors or parity-violating gauge fields \cite{Watanabe2009,Soda2012Statistical,Maleknejad2013Review}) and asking whether any reported large-scale anomalies can be reproduced without violating isotropy constraints at smaller scales. More generally, whether the EC ekpyrotic phase actually tames Belinski--Khalatnikov--Lifshitz (BKL)--type mixmaster dynamics in inhomogeneous collapse remains an open question; here I only control the homogeneous (Bianchi-type) sector.

    On the microphysical side, the Weyssenhoff parameter $\alpha$ should be tied more explicitly to the underlying matter content, so that astrophysical and laboratory constraints on spin density can be translated into bounds on the bounce scale. Finally, if one were to embed this kind of Einstein--Cartan bounce in a cyclic cosmology, any realistic scenario would have to track the entropy budget across cycles, including black-hole channels, particle production, and possible dilution or reset mechanisms. Addressing these questions would connect the background construction developed here to concrete observational tests and clarify under what conditions Einstein--Cartan based bouncing cosmologies can be treated as mathematically self-consistent effective background alternatives to $\Lambda$CDM, with observational viability to be assessed once perturbations and data comparisons are carried out.

    \clearpage
    \appendix
    \section{Comparison with Copeland--Liddle--Wands (1998)}
    \label{app:clw_comparison}
    As a consistency check on both the extended system and the numerical implementation, I first reproduced the Copeland--Liddle--Wands (CLW) system in the spatially flat, isotropic, torsionless limit \cite{Copeland1998Scaling}. Setting $\Sigma=\Omega_k=\Omega_s=0$ and $s=+1$ in Eq.~\eqref{eq:Friedmann} gives
    \begin{equation}
        x=\frac{\dot\phi}{\sqrt{6}\,H},\qquad
        y=\frac{\sqrt{V(\phi)}}{\sqrt{3}\,H},\qquad
        z=\frac{\rho_m}{3H^2},\qquad
        x^2+y^2+z=1,
    \end{equation}
    and the $N=\ln a$ equations reduce to
    \begin{align}
        x' &= -3x + \sqrt{\tfrac{3}{2}}\,\lambda y^2 
        + \tfrac{3}{2}x\,(2x^2+\gamma_m z),\\
        y' &= -\sqrt{\tfrac{3}{2}}\,\lambda xy 
        + \tfrac{3}{2}y\,(2x^2+\gamma_m z),\\
        z' &= -3\gamma_m z 
        + 3z\,(2x^2+\gamma_m z),
    \end{align}
    matching the CLW system on the Friedmann constraint surface $x^2+y^2+z=1$, since the $2x^2+\gamma_m z$ friction term is equivalent to the usual $2x^2+\gamma_m(1-x^2-y^2)$ once the constraint is imposed. In this limit the fixed points and their stability properties coincide with the original CLW analysis. The scalar-field dominated point is stable for $\lambda^2<3\gamma_m$ and drives accelerated expansion when $\lambda^2<2$; the scaling (tracking) solution exists and is stable for $\lambda^2>3\gamma_m$; and the kinetic $\pm$ points are always unstable. Evaluating the $6\times6$ Jacobian in the  extended system and then restricting to $(x,y,z)$ reproduces the CLW eigenvalues; in the CLW limit $(\Sigma,\Omega_k,\Omega_s)=0$ the full $6\times6$ Jacobian is block-upper-triangular, so the extra geometric directions do not affect the $(x,y,z)$ eigenvalues. Phase portraits and time evolution in this limit (not shown) agree with the standard behavior, confirming that the extended EC model reduces consistently to CLW in the flat, isotropic, torsionless case and that the numerics are benchmarked against this baseline.

    \section{From CLW to the extended EC + shear + curvature system and the compact $q$-form}
    \label{app:derivation}

    \textbf{CLW baseline and extension.}
    The CLW variables $(x,y,z)$, the sign parameter $s\equiv\mathrm{sign}(V)$, and $w_\phi(x,y)$ are defined in Sec.~\ref{sec:assumptions}, and the flat, isotropic GR limit reproduces the CLW system and fixed points (Appendix~\ref{app:clw_comparison}). To extend this to the Einstein--Cartan + shear + curvature setting, I add the normalized geometric variables
    \[
        \Sigma \equiv \frac{\sigma}{H},\qquad
        \Omega_k \equiv -\frac{k}{a^2H^2},\qquad
        \Omega_s \equiv \frac{\rho_s}{3H^2},\qquad
        \rho_s\propto a^{-6},
    \]
    and replace the flat GR Friedmann constraint by Eq.~\eqref{eq:Friedmann}.

    \textbf{Deceleration parameter $q$ and the canonical identity.}
    The deceleration parameter is defined by
    \begin{equation}
        q \equiv -1 - \frac{\dot{H}}{H^2}.
    \end{equation}
    Using the EC-modified Raychaudhuri equation and the normalized variables, the total pressure
    \begin{equation}
        p_{\rm tot} = p_\phi + p_m
        = 3H^2(x^2 - s y^2) + 3H^2(\gamma_m -1)z
    \end{equation}
    and the anisotropy, curvature, and spin contributions lead to
    \begin{equation}
        q = \tfrac{1}{2}\Big[
            1 + 3(x^2 - s y^2)
            + 3(\gamma_m - 1)z
            + 3\Sigma^2
            - \Omega_k
            - 3\Omega_s \Big].
    \end{equation}
    Eliminating $s y^2$ with the extended Friedmann constraint,
    \begin{equation}
        s y^2 = 1 - x^2 - z - \Sigma^2 - \Omega_k + \Omega_s,
    \end{equation}
    gives the compact canonical form
    \begin{equation}
        \boxed{
        q = 3x^2 - 1
            + \tfrac{3}{2}\gamma_m z
            + 3\Sigma^2
            + \Omega_k
            - 3\Omega_s}.
    \end{equation}
    This coincides with the canonical expression~\eqref{eq:q_canonical} used in the main text.

    \textbf{Explicit extended system.}
    Starting from the scalar and fluid continuity equations,
    \[
        \dot\rho_\phi + 3H(\rho_\phi + p_\phi)=0,\qquad
        \dot\rho_m + 3\gamma_m H\rho_m=0,
    \]
    and using $d/dN = H^{-1}d/dt$ and the definitions of $(x,y,z)$, one recovers the CLW system in the flat, isotropic GR limit. In the extended EC setting, one simply rewrites the CLW "friction term" $ \tfrac{3}{2}(2x^2+\gamma_m z)$ as $1+q$ using Eq.~\eqref{eq:q_canonical}, which gives the explicit evolution equations
    \begin{align}
        x' &= -3x + s\sqrt{\tfrac{3}{2}}\,\lambda y^2 + x(1+q), \label{eq:ext-x}\\
        y' &= -\sqrt{\tfrac{3}{2}}\,\lambda xy + y(1+q), \label{eq:ext-y}\\
        z' &= z\bigl[-3\gamma_m + 2(1+q)\bigr]. \label{eq:ext-z}
    \end{align}
    In the GR, flat, isotropic limit $(\Sigma,\Omega_k,\Omega_s)\to 0$, substituting Eq.~\eqref{eq:q_canonical} into
    Eqs.~\eqref{eq:ext-x}--\eqref{eq:ext-z} reproduces the original CLW equations exactly.

    \textbf{Geometry-sector evolution.}
    The shear and curvature variables obey the standard evolution equations \cite{WainwrightEllis1997}, and for a Weyssenhoff spin fluid with $\rho_s \propto a^{-6}$ the normalized spin density evolves analogously. Collecting these, the geometry-sector variables satisfy
    \begin{equation}
        \boxed{
        \Sigma' = (q-2)\Sigma,\qquad
        \Omega_k' = 2q\,\Omega_k,\qquad
        \Omega_s' = 2(q-2)\Omega_s
        }.
        \label{eq:geom-evol}
    \end{equation}

    \textbf{Ekpyrotic shear suppression.}
    In the ekpyrotic regime the potential is steep and negative ($V<0$, $s=-1$), and for an exponential potential
    $V(\phi) = -V_0 e^{-\lambda\phi}$ with constant $\lambda$, the scalar approaches a scaling solution with
    \begin{equation}
        w_\phi = \frac{\lambda^2}{3} - 1 > 1.
    \end{equation}
    When the scalar dominates, $w_{\rm tot} \simeq w_\phi$ and
    \begin{equation}
        q \simeq \frac{1 + 3w_\phi}{2} > 2.
    \end{equation}
    Using Eq.~\eqref{eq:geom-evol}, the shear evolves as
    \begin{equation}
        \Sigma' = (q-2)\Sigma
                = \Big(\frac{1 + 3w_\phi}{2} - 2 \Big)\Sigma
                = \frac{3}{2}(w_\phi - 1)\Sigma.
    \end{equation}
    Defining $\mu_\Sigma \equiv \tfrac{3}{2}(w_\phi - 1) > 0$, and recalling that on the contracting branch $N\to -\infty$, 
    \begin{equation}
        \Sigma(N) \propto e^{\mu_\Sigma N}
        \xrightarrow{N \to -\infty} 0.
    \end{equation}
    Thus a steep negative exponential potential generates an ekpyrotic attractor that exponentially suppresses shear anisotropy, so homogeneous shear decays during the contracting phase within the extended Einstein--Cartan framework.

    \section{Jacobian derivation for the extended system}
    \label{app:ecem_jacobian}
    I now derive the Jacobian of the autonomous system for a canonical scalar field with potential $V(\phi)$ in the extended Einstein--Cartan plus shear/curvature setting. For the linear stability analysis, $\lambda$ is treated as constant (as in an exponential potential, or the local value of $\lambda(\phi)$ at a fixed point). The sign parameter $s \equiv \text{sign}(V)$ enters only in the $x'$ equation; this choice preserves the Friedmann constraint (see Appendix~\ref{app:derivation}).

    The evolution equations for $(x,y,z,\Sigma,\Omega_k,\Omega_s)$ are given by Eqs.~\eqref{eq:ext-x}--\eqref{eq:ext-z} together with the geometric sector in Eq.~\eqref{eq:geom-evol}, with the canonical deceleration parameter in Eq.~\eqref{eq:q_canonical}. These are equivalent to Eqs.~\eqref{eq:dyn_xy}--\eqref{eq:dyn_ks} in the main text; I retain the compact $q$-form because it makes the structure of the dynamics and the stability analysis more transparent.

    \subsection{Partial derivatives of $q$}
    From the canonical expression above,
    \begin{align}
        \frac{\partial q}{\partial x} &= 6x, &
        \frac{\partial q}{\partial y} &= 0, &
        \frac{\partial q}{\partial z} &= \tfrac{3}{2}\gamma_m, \\
        \frac{\partial q}{\partial \Sigma} &= 6\Sigma, &
        \frac{\partial q}{\partial \Omega_k} &= 1, &
        \frac{\partial q}{\partial \Omega_s} &= -3.
    \end{align}
    The derivative $\partial q / \partial y$ vanishes because $y$ has been eliminated from $q$ using the Friedmann constraint (Appendix~\ref{app:derivation}).

    \subsection{Jacobian entries}
    Define
    \begin{equation}
        J \equiv \frac{\partial(f_x,f_y,f_z,f_\Sigma,f_k,f_s)}
                    {\partial(x,y,z,\Sigma,\Omega_k,\Omega_s)},
    \end{equation}
    where $f_x,\dots,f_s$ denote the six right-hand sides and $f_k \equiv \Omega_k'$, $f_s \equiv \Omega_s'$.

    \smallskip
    \noindent\textbf{Row 1: derivatives of $f_x$.}\\
    With $f_x = -3x + s\sqrt{\tfrac{3}{2}}\,\lambda y^2 + x(1+q)$,
    \begin{align}
        \frac{\partial f_x}{\partial x} &= -3 + (1+q) + x\frac{\partial q}{\partial x} = q - 2 + 6x^2,\\
        \frac{\partial f_x}{\partial y} &= 2s\sqrt{\tfrac{3}{2}}\,\lambda y + x\frac{\partial q}{\partial y} = \sqrt{6}\,s\lambda y,\\
        \frac{\partial f_x}{\partial z} &= x\frac{\partial q}{\partial z} = \tfrac{3}{2}\gamma_m x,\\
        \frac{\partial f_x}{\partial \Sigma} &= x\frac{\partial q}{\partial \Sigma} = 6x\Sigma,\\
        \frac{\partial f_x}{\partial \Omega_k} &= x\frac{\partial q}{\partial \Omega_k} = x,\\
        \frac{\partial f_x}{\partial \Omega_s} &= x\frac{\partial q}{\partial \Omega_s} = -3x.
    \end{align}

    \smallskip
    \noindent\textbf{Row 2: derivatives of $f_y$.}\\
    With $f_y = -\sqrt{\tfrac{3}{2}}\,\lambda xy + y(1+q)$,
    \begin{align}
        \frac{\partial f_y}{\partial x} &= -\sqrt{\tfrac{3}{2}}\,\lambda y + y\frac{\partial q}{\partial x} = y\Big(6x - \sqrt{\tfrac{3}{2}}\lambda\Big),\\
        \frac{\partial f_y}{\partial y} &= -\sqrt{\tfrac{3}{2}}\,\lambda x + (1+q),\\
        \frac{\partial f_y}{\partial z} &= y\frac{\partial q}{\partial z} = \tfrac{3}{2}\gamma_m y,\\
        \frac{\partial f_y}{\partial \Sigma} &= y\frac{\partial q}{\partial \Sigma} = 6y\Sigma,\\
        \frac{\partial f_y}{\partial \Omega_k} &= y\frac{\partial q}{\partial \Omega_k} = y,\\
        \frac{\partial f_y}{\partial \Omega_s} &= y\frac{\partial q}{\partial \Omega_s} = -3y.
    \end{align}

    \smallskip
    \noindent\textbf{Row 3: derivatives of $f_z$.}\\
    With $f_z = z[-3\gamma_m + 2(1+q)]$,
    \begin{align}
        \frac{\partial f_z}{\partial x} &= z\cdot 2\frac{\partial q}{\partial x} = 2z(6x) = 12xz,\\
        \frac{\partial f_z}{\partial y} &= 2z\frac{\partial q}{\partial y} = 0,\\
        \frac{\partial f_z}{\partial z} &= [-3\gamma_m + 2(1+q)] + z\cdot 2\frac{\partial q}{\partial z} = -3\gamma_m + 2(1+q) + 3\gamma_m z,\\
        \frac{\partial f_z}{\partial \Sigma} &= 2z\frac{\partial q}{\partial \Sigma} = 12z\Sigma,\\
        \frac{\partial f_z}{\partial \Omega_k} &= 2z\frac{\partial q}{\partial \Omega_k} = 2z,\\
        \frac{\partial f_z}{\partial \Omega_s} &= 2z\frac{\partial q}{\partial \Omega_s} = -6z.
    \end{align}

    \smallskip
    \noindent\textbf{Row 4: derivatives of $f_\Sigma$.}\\
    With $f_\Sigma = -(2-q)\Sigma = (q-2)\Sigma$,
    \begin{align}
        \frac{\partial f_\Sigma}{\partial x} &= \Sigma\frac{\partial q}{\partial x} = 6x\Sigma,\\
        \frac{\partial f_\Sigma}{\partial y} &= 0,\\
        \frac{\partial f_\Sigma}{\partial z} &= \Sigma\frac{\partial q}{\partial z} = \tfrac{3}{2}\gamma_m \Sigma,\\
        \frac{\partial f_\Sigma}{\partial \Sigma} &= (q-2) + \Sigma\frac{\partial q}{\partial \Sigma} = q - 2 + 6\Sigma^2,\\
        \frac{\partial f_\Sigma}{\partial \Omega_k} &= \Sigma\frac{\partial q}{\partial \Omega_k} = \Sigma,\\
        \frac{\partial f_\Sigma}{\partial \Omega_s} &= \Sigma\frac{\partial q}{\partial \Omega_s} = -3\Sigma.
    \end{align}

    \smallskip
    \noindent\textbf{Row 5: derivatives of $f_k$.}\\
    With $f_k = 2q\Omega_k$,
    \begin{align}
        \frac{\partial f_k}{\partial x} &= 2\Omega_k\frac{\partial q}{\partial x} = 12x\Omega_k,\\
        \frac{\partial f_k}{\partial y} &= 0,\\
        \frac{\partial f_k}{\partial z} &= 2\Omega_k\frac{\partial q}{\partial z} = 3\gamma_m \Omega_k,\\
        \frac{\partial f_k}{\partial \Sigma} &= 2\Omega_k\frac{\partial q}{\partial \Sigma} = 12\Sigma\Omega_k,\\
        \frac{\partial f_k}{\partial \Omega_k} &= 2q + 2\Omega_k\frac{\partial q}{\partial \Omega_k} = 2q + 2\Omega_k,\\
        \frac{\partial f_k}{\partial \Omega_s} &= 2\Omega_k\frac{\partial q}{\partial \Omega_s} = -6\Omega_k.
    \end{align}

    \smallskip
    \noindent\textbf{Row 6: derivatives of $f_s$.}\\
    With $f_s = 2(q-2)\Omega_s$,
    \begin{align}
        \frac{\partial f_s}{\partial x} &= 2\Omega_s\frac{\partial q}{\partial x} = 12x\Omega_s,\\
        \frac{\partial f_s}{\partial y} &= 0,\\
        \frac{\partial f_s}{\partial z} &= 2\Omega_s\frac{\partial q}{\partial z} = 3\gamma_m \Omega_s,\\
        \frac{\partial f_s}{\partial \Sigma} &= 2\Omega_s\frac{\partial q}{\partial \Sigma} = 12\Sigma\Omega_s,\\
        \frac{\partial f_s}{\partial \Omega_k} &= 2\Omega_s\frac{\partial q}{\partial \Omega_k} = 2\Omega_s,\\
        \frac{\partial f_s}{\partial \Omega_s} &= 2(q-2) + 2\Omega_s\frac{\partial q}{\partial \Omega_s} = 2(q-2) - 6\Omega_s.
    \end{align}

    \subsection{Final Jacobian (canonical form)}

    Collecting all entries, the Jacobian in the canonical variables
    $(x,y,z,\Sigma,\Omega_k,\Omega_s)$ is
    \begin{widetext}
    \begingroup
    \setlength{\arraycolsep}{4pt}
    \begin{equation*}
        \resizebox{0.98\textwidth}{!}{$\displaystyle
        \begin{pmatrix}
        q - 2 + 6x^{2}              & \sqrt{6}\,s\lambda y                          & \tfrac{3}{2}\gamma_m x                  & 6x\Sigma                 & x                 & -3x \\
        y\!\left(6x - \sqrt{\tfrac{3}{2}}\lambda\right) & 1 + q - \sqrt{\tfrac{3}{2}}\lambda x & \tfrac{3}{2}\gamma_m y                  & 6y\Sigma                 & y                 & -3y \\
        12xz                        & 0                                           & -3\gamma_m + 2(1+q) + 3\gamma_m z       & 12z\Sigma                & 2z                & -6z \\
        6x\Sigma                    & 0                                           & \tfrac{3}{2}\gamma_m \Sigma             & q - 2 + 6\Sigma^{2}      & \Sigma            & -3\Sigma \\
        12x\Omega_k                 & 0                                           & 3\gamma_m \Omega_k                      & 12\Sigma\Omega_k         & 2q + 2\Omega_k    & -6\Omega_k \\
        12x\Omega_s                 & 0                                           & 3\gamma_m \Omega_s                      & 12\Sigma\Omega_s         & 2\Omega_s         & 2(q-2) - 6\Omega_s
        \end{pmatrix}
        $}.
    \end{equation*}
    \endgroup
    \end{widetext}
    This is the form used in Sec.~\ref{sec:eigenvalues} to evaluate the eigenvalues at each fixed point, with $q$ given in terms of $(x,y,z,\Sigma,\Omega_k,\Omega_s)$ by Eq.~\eqref{eq:q_canonical}.

    \section{Kinetic dominated fixed points: full Jacobian evaluation}
    \label{app:eigen_derivations}
    The kinetic-dominated fixed points listed in Sec.~\ref{sec:eigenvalues} have
    \(
        (x_*,y_*,z_*,\Sigma_*,\Omega_{k*},\Omega_{s*})=(\pm1,0,0,0,0,0)
    \)
    and $q_*=2$ from Eq.~\eqref{eq:q_canonical}.

    \subsection{Jacobian evaluation}
    Using the general Jacobian $J$ of Appendix~\ref{app:ecem_jacobian} and substituting $x_*=\pm1$ with
    \[
        y_*=z_*=\Sigma_*=\Omega_{k*}=\Omega_{s*}=0,
    \]
    all lower off-diagonal terms vanish because they carry factors of $y,z,\Sigma,\Omega_k,\Omega_s$. The resulting matrix is \emph{upper triangular}, so its eigenvalues are just the diagonal entries:
    \begin{align}
        \mu_x &= \left.\frac{\partial f_x}{\partial x}\right|_* = q_* - 2 + 6x_*^2 
            = 2 - 2 + 6(1) = 6,\\[4pt]
        \mu_y &= \left.\frac{\partial f_y}{\partial y}\right|_* = (1+q_*) - \sqrt{\tfrac{3}{2}}\lambda x_* 
            = 3 \mp \sqrt{\tfrac{3}{2}}\lambda,\\[4pt]
        \mu_z &= \left.\frac{\partial f_z}{\partial z}\right|_* = -3\gamma_m + 2(1+q_*) 
            = -3\gamma_m + 6 = 3(2-\gamma_m),\\[4pt]
        \mu_\Sigma &= \left.\frac{\partial f_\Sigma}{\partial \Sigma}\right|_* = q_* - 2 = 2 - 2 = 0,\\[4pt]
        \mu_k &= \left.\frac{\partial f_k}{\partial \Omega_k}\right|_* = 2q_* = 4,\\[4pt]
        \mu_s &= \left.\frac{\partial f_s}{\partial \Omega_s}\right|_* = 2(q_*-2) = 0.
    \end{align}
    Here the plus sign in $\mu_y = 3 \mp \sqrt{\tfrac{3}{2}}\lambda$ corresponds to $x_* = +1$, and the minus sign to $x_* = -1$.

    \subsection{Spectrum and stability interpretation}
    The spectrum can be written compactly as
    \begin{equation}
        \{\mu\} = \Big\{ 6,\;\; 3 \mp \sqrt{\tfrac{3}{2}}\,\lambda,\;\; 3(2-\gamma_m),\;\; 0,\;\; 4,\;\; 0 \Big\}.
    \end{equation}

    The zero eigenvalues for shear and spin--torsion,
    \begin{equation}
        \mu_\Sigma = 0,\qquad \mu_s = 0,
    \end{equation}
    show that the kinetic fixed points ($w_\phi = 1$) are only \emph{marginally} stable in these directions: a pure stiff fluid does not dynamically suppress anisotropy or spin relative to the background energy density, so the homogeneous shear instability is not cured by the stiff phase alone.

    This marginality matches the equal scaling of the shear and kinetic components in a stiff phase: $\rho_\sigma \propto a^{-6}$ and $\rho_\phi \propto a^{-6}$ for $w_\phi=1$, so their ratio remains constant. By contrast, in the ekpyrotic regime with $w_\phi>1$ the scalar redshifts faster than $a^{-6}$ and the shear fraction is dynamically driven to zero (Sec.~\ref{sec:shear_suppression}).

    The physically relevant ekpyrotic regime corresponds not to the exact kinetic point, but to nearby trajectories driven by a steep negative potential with $w_\phi > 1$. In that regime the effective shear eigenvalue satisfies
    \begin{equation}
        \mu_\Sigma^{\rm eff} \approx \frac{3}{2}(w_\phi - 1) > 0,
    \end{equation}
    so on the contracting branch ($N\to -\infty$) one has $\Sigma \propto e^{\mu_\Sigma^{\rm eff} N} \to 0$. Thus the ekpyrotic phase behaves as a genuine anisotropy attractor even though the limiting stiff fixed point itself is only marginal in the shear and spin directions.

    \section{Fluid-dominated fixed point: full Jacobian evaluation}
    \label{app:ecem_fluid_eig}

    The fluid fixed point in Sec.~\ref{sec:eigenvalues} has
    \begin{equation}
        (x_*,y_*,z_*,\Sigma_*,\Omega_{k*},\Omega_{s*}) = (0,0,1,0,0,0),
    \end{equation}
    corresponding to a pure barotropic fluid with $w_m=\gamma_m-1$. From the canonical deceleration parameter \eqref{eq:q_canonical}:
    \begin{equation}
        q_* = \frac{1}{2}\big(1+3w_m\big)
            = \frac{3\gamma_m-2}{2},
        \qquad
        1+q_* = \frac{3\gamma_m}{2}.
    \end{equation}

    Inserting this fixed point into the general Jacobian of Appendix~\ref{app:ecem_jacobian}, and noting that all entries proportional to $x,y,\Sigma,\Omega_k,\Omega_s$ vanish, the Jacobian becomes upper triangular, so the eigenvalues are given by the diagonal entries. The non-zero diagonal entries are
    \begin{align}
        \mu_x &= \left.\frac{\partial f_x}{\partial x}\right|_*
            = q_* - 2
            = \frac{3}{2}(\gamma_m-2),\\[2pt]
        \mu_y &= \left.\frac{\partial f_y}{\partial y}\right|_*
            = 1+q_*
            = \frac{3}{2}\gamma_m,\\[2pt]
        \mu_z &= \left.\frac{\partial f_z}{\partial z}\right|_*
            = -3\gamma_m + 2(1+q_*) + 3\gamma_m z_*
            = 3\gamma_m,\\[2pt]
        \mu_\Sigma &= \left.\frac{\partial f_\Sigma}{\partial \Sigma}\right|_*
            = q_* - 2
            = \frac{3}{2}(\gamma_m-2),\\[2pt]
        \mu_k &= \left.\frac{\partial f_k}{\partial \Omega_k}\right|_*
            = 2q_*
            = 3\gamma_m-2,\\[2pt]
        \mu_s &= \left.\frac{\partial f_s}{\partial \Omega_s}\right|_*
            = 2(q_*-2)
            = 3(\gamma_m-2).
    \end{align}

    Collecting these, the full spectrum at the fluid point is
    \begin{equation}
        \boxed{
        \begin{aligned}
        \{\mu\} =
        \Big\{&
            \tfrac{3}{2}(\gamma_m-2),\;
            \tfrac{3}{2}\gamma_m,\;
            3\gamma_m,\\
            &
            \tfrac{3}{2}(\gamma_m-2),\;
            3\gamma_m-2,\;
            3(\gamma_m-2)
        \Big\},
        \end{aligned}
        }
    \end{equation}
    in agreement with Table~\ref{tab:extended_fp}.
        
    \section{Scalar-field dominated fixed point: full Jacobian evaluation}
    \label{app:ecem_scalar_eig}

    \subsection{Coordinates and kinematics}
    The scalar-field dominated fixed point listed in Sec.~\ref{sec:eigenvalues} (late-time quintessence with $V>0$) exists for $\lambda^2<6$ and has
    \begin{equation}
        (x_*,y_*,z_*,\Sigma_*,\Omega_{k*},\Omega_{s*})
        =
        \Big(\tfrac{\lambda}{\sqrt{6}},\sqrt{1-\tfrac{\lambda^2}{6}},0,0,0,0\Big).
    \end{equation}
    From Eq.~\eqref{eq:q_canonical},
    \begin{equation}
        q_* = \frac{\lambda^2}{2} - 1,\qquad 1+q_* = \frac{\lambda^2}{2},
    \end{equation}
    and the associated scalar equation of state is $w_\phi = \lambda^2/3 - 1$ (accelerated for $\lambda^2<2$). Substituting this fixed point into the general Jacobian of Appendix~\ref{app:ecem_jacobian} gives the eigenvalues quoted in Sec.~\ref{sec:eigenvalues}; for completeness I sketch the evaluation.

    \subsection{Jacobian at the fixed point}
    Starting from the canonical Jacobian in Appendix~\ref{app:ecem_jacobian}, evaluate $J$ at $(x_*,y_*,z_*,\Sigma_*,\Omega_{k*},\Omega_{s*})$.

    \paragraph*{The $(x,y)$ block.}
    From Appendix~\ref{app:ecem_jacobian},
    \begin{align}
        \left.\frac{\partial f_x}{\partial x}\right|_* &= q_* - 2 + 6x_*^2
            = \frac{3}{2}\lambda^2 - 3,\\[4pt]
        \left.\frac{\partial f_x}{\partial y}\right|_* &= \sqrt{6}\,s\lambda y_*
            = \lambda\sqrt{6-\lambda^2},\\[4pt]
        \left.\frac{\partial f_y}{\partial x}\right|_* &= y_*\left(6x_* - \sqrt{\tfrac{3}{2}}\lambda\right)
            = \frac{\lambda}{2}\sqrt{6-\lambda^2},\\[4pt]
        \left.\frac{\partial f_y}{\partial y}\right|_* &= (1+q_*) - \sqrt{\tfrac{3}{2}}\lambda x_*
            = \frac{\lambda^2}{2} - \frac{\lambda^2}{2} = 0,
    \end{align}
    where $s=+1$ and $x_*=\lambda/\sqrt{6}$ are used. Thus
    \begin{equation}
        J_{xy}\Big|_* =
        \begin{bmatrix}
            \tfrac{3}{2}\lambda^2 - 3 & \lambda\sqrt{6-\lambda^2} \\
            \tfrac{\lambda}{2}\sqrt{6-\lambda^2} & 0
        \end{bmatrix}.
    \end{equation}
    The eigenvalues of this $2\times 2$ block follow from
    \begin{equation}
        \det\big(J_{xy}|_* - \mu I\big)
        = \frac{1}{2}\big(\lambda^2 - \mu\big)\big(\lambda^2 - 2\mu - 6\big)=0,
    \end{equation}
    so
    \begin{equation}
        \mu_{xy}^{(1)} = \lambda^2,\qquad
        \mu_{xy}^{(2)} = \frac{\lambda^2 - 6}{2}.
    \end{equation}
    The $\mu=\lambda^2$ direction is normal to the Friedmann constraint surface; on the physical $(x,y,z)$ manifold with $x^2 + y^2 + z = 1$, the remaining scalar-sector eigenvalue is $\mu_{xy}^{(2)}$.

    \paragraph*{Matter $z$.}
    From Appendix~\ref{app:ecem_jacobian},
    \begin{align}
        \left.\frac{\partial f_z}{\partial z}\right|_*
            &= -3\gamma_m + 2(1+q_*) \notag\\
            &= -3\gamma_m + \lambda^2
            \quad\Rightarrow\quad
        \mu_z = \lambda^2 - 3\gamma_m.
    \end{align}

    \paragraph*{Shear $\Sigma$.}
    At the fixed point,
    \begin{equation}
        \mu_\Sigma
        = \left.\frac{\partial f_\Sigma}{\partial \Sigma}\right|_*
        = q_* - 2
        = \frac{\lambda^2}{2} - 3
        = \frac{\lambda^2 - 6}{2}.
    \end{equation}

    \paragraph*{Curvature $\Omega_k$.}
    For the curvature mode,
    \begin{equation}
        \mu_k
        = \left.\frac{\partial f_k}{\partial \Omega_k}\right|_*
        = 2q_* = \lambda^2 - 2.
    \end{equation}

    \paragraph*{Spin--torsion $\Omega_s$.}
    Finally, for the Weyssenhoff spin sector,
    \begin{equation}
        \mu_s
        = \left.\frac{\partial f_s}{\partial \Omega_s}\right|_*
        = 2(q_* - 2) = \lambda^2 - 6.
    \end{equation}

    \subsection{Spectrum and stability}
    Collecting all six eigenvalues of the full $(x,y,z,\Sigma,\Omega_k,\Omega_s)$ system,
    \begin{equation}
        \boxed{
            \{\mu\} =
            \Big\{\lambda^2,\;
                \tfrac{\lambda^2 - 6}{2},\;
                \lambda^2 - 3\gamma_m,\;
                \tfrac{\lambda^2 - 6}{2},\;
                \lambda^2 - 2,\;
                \lambda^2 - 6
            \Big\}.
        }
    \end{equation}
    The eigenvalue $\mu=\lambda^2$ corresponds to motion off the Friedmann constraint surface and is removed when one restricts to the physical phase space. On the constrained manifold, the scalar+fluid sector reduces to the familiar CLW eigenvalues $\{(\lambda^2 - 6)/2,\ \lambda^2 - 3\gamma_m\}$, while the extended EC + shear + curvature system adds the geometric eigenvalues $\mu_\Sigma=(\lambda^2-6)/2$, $\mu_k=\lambda^2-2$, and $\mu_s=\lambda^2-6$.

    \textbf{Sector stability.}
    Requiring all physical directions to be stable (negative eigenvalues) yields:
    \begin{equation}
        \begin{aligned}
            &\mu_{xy}^{(2)}<0 \;\Leftrightarrow\; \lambda^2<6 \quad &(\text{scalar sector}),\\
            &\mu_z<0 \;\Leftrightarrow\; \lambda^2<3\gamma_m \quad &(\text{matter}),\\
            &\mu_\Sigma<0 \;\Leftrightarrow\; \lambda^2<6 \quad &(\text{shear}),\\
            &\mu_k<0 \;\Leftrightarrow\; \lambda^2<2 \quad &(\text{curvature}),\\
            &\mu_s<0 \;\Leftrightarrow\; \lambda^2<6 \quad &(\text{spin--torsion}).
        \end{aligned}
    \end{equation}
    Thus the scalar-field dominated point is a late-time attractor with decaying matter, shear, torsion and curvature provided $\lambda^2<3\gamma_m$ and, in particular, $\lambda^2<2$ for curvature stability and accelerated expansion.

    \section{Scaling fixed point: eigenvalues with anisotropy, curvature, and spin}
    \label{app:ecem_scaling_eig}
    \subsection{Coordinates, existence, and kinematics}
    The scaling (tracking) fixed point in the extended system is summarized in Sec.~\ref{sec:eigenvalues}. It has
    \begin{equation}
        \begin{split}
        &(x_*,y_*,z_*,\Sigma_*,\Omega_{k*},\Omega_{s*})\\
        &\quad =
        \left(
            \tfrac{\sqrt{6}\gamma_m}{2\lambda},\,
            \sqrt{\tfrac{3\gamma_m(2-\gamma_m)}{2\lambda^2}},\,
            1-\tfrac{3\gamma_m}{\lambda^2},\,
            0,0,0
        \right).
        \end{split}
    \end{equation}
    and exists for $0<\gamma_m<2$ and $\lambda^2>3\gamma_m$, where the scalar tracks the background fluid ($w_\phi\to w_m=\gamma_m-1$ and $w_{\rm eff}=w_m$). Inserting these coordinates into Eq.~\eqref{eq:q_canonical} gives
    \begin{equation}
        1+q_* = \frac{3\gamma_m}{2},\qquad q_* = \frac{3\gamma_m}{2} - 1,
    \end{equation}
    consistent with $q_*=\tfrac{1}{2}(1+3w_m)$. Substituting this fixed point into the general Jacobian of Appendix~\ref{app:ecem_jacobian} yields the eigenvalues $\mu_\pm,\mu_\Sigma,\mu_k,\mu_s$ quoted in Sec.~\ref{sec:eigenvalues}; the explicit forms are recorded here for completeness.

    \subsection{Eigenvalues}
    \paragraph*{(i) Scalar--matter sector (CLW block).}
    On the Friedmann constraint surface, the $(x,y,z)$ sector reduces to the standard CLW scaling solution. The two non-trivial eigenvalues in the scalar--matter subspace are \cite{Copeland1998Scaling}
    \begin{equation}
        \boxed{
        \mu_{\pm}
        = -\frac{3(2-\gamma_m)}{4}
        \left[
            1 \pm \sqrt{
                1 - \frac{8\gamma_m(\lambda^2-3\gamma_m)}{\lambda^2(2-\gamma_m)}
            }
        \right]
        }
    \end{equation}
    (valid for $\lambda^2>3\gamma_m$, $0<\gamma_m<2$). For $0<\gamma_m<2$ and $\lambda^2>3\gamma_m$, both $\mu_\pm<0$, so the CLW block is attractive in the flat, shearless, torsionless subspace.

    \paragraph*{(ii) Shear, curvature, and spin--torsion.}
    The geometric sectors follow directly from
    \begin{equation}
        \Sigma'=(q-2)\Sigma,\qquad
        \Omega_k'=2q\,\Omega_k,\qquad
        \Omega_s'=2(q-2)\Omega_s,
    \end{equation}
    evaluated at $q_*=\tfrac{3\gamma_m}{2}-1$:
    \begin{equation}
        \boxed{
        \mu_\Sigma=\frac{3}{2}(\gamma_m-2),\qquad
        \mu_k=3\gamma_m-2,\qquad
        \mu_s=3(\gamma_m-2).
        }
    \end{equation}
    Thus shear and spin are non-growing (indeed decaying for $\gamma_m<2$), while curvature has a distinct eigenvalue $\mu_k=3\gamma_m-2$ that becomes positive for standard fluids ($\gamma_m\ge1$).

    \subsection{Summary and stability properties}
    The eigenvalues at all fixed points are summarized in Table~\ref{tab:extended_fp}. For the scaling point, the key results are:
    \begin{itemize}
        \item \textbf{CLW block:} $\mu_\pm<0$ for $\lambda^2>3\gamma_m$ and $0<\gamma_m<2$; the point is attractive in the
            $(x,y,z)$ subspace (node or spiral depending on the discriminant).
        \item \textbf{Shear:} $\mu_\Sigma = \tfrac{3}{2}(\gamma_m-2)\le 0$ for $\gamma_m\le2$; shear decays in the physical range $0<\gamma_m<2$.
        \item \textbf{Spin--torsion:} $\mu_s = 3(\gamma_m-2)\le 0$ for $\gamma_m\le2$; spin density is also damped.
        \item \textbf{Curvature:} $\mu_k = 3\gamma_m-2$. For dust and radiation ($\gamma_m\ge1$), one has $\mu_k>0$ so curvature is a growing (relevant) direction.
    \end{itemize}
    Consequently, even when the CLW block is fully attractive, the scaling point is a saddle in the full six-dimensional $(x,y,z,\Sigma,\Omega_k,\Omega_s)$ space once curvature is allowed to evolve.

    \section{Additional numerical simulation results}
    \label{app:ecem_sims}
\begin{figure}[htpb]
        \centering
        \includegraphics[width=0.75\linewidth]{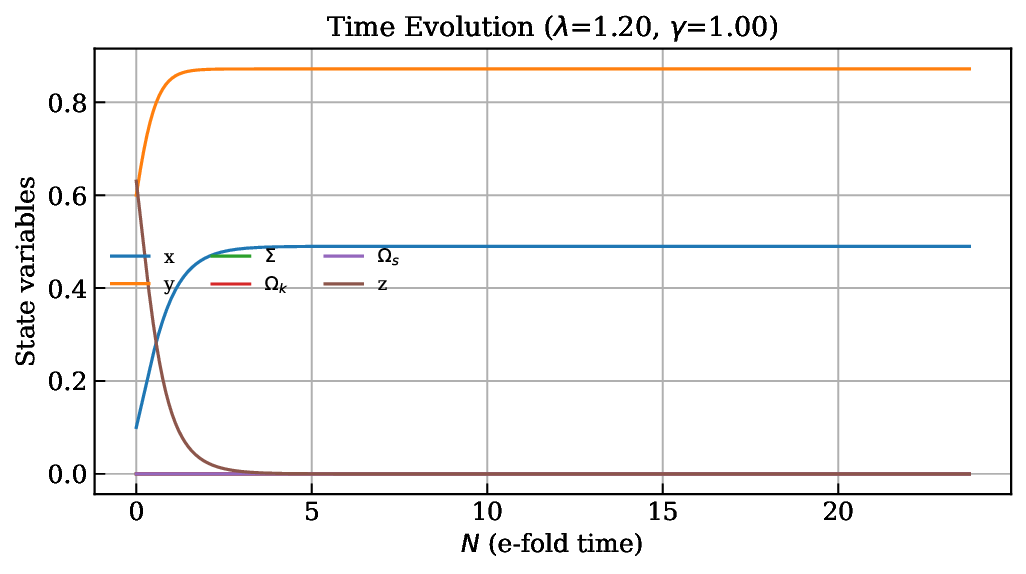}
        \caption{Set (a), $(\lambda,\gamma_m)=(1.20,1.00)$. Time evolution of the normalized variables and effective equation of state for the trajectory shown in Fig.~\ref{fig:ecem_intro_phase}(a). The orbit approaches the scalar-field dominated node $(x_*,y_*)\approx(0.49,0.87)$, while $z,\Omega_k,\Sigma,\Omega_s$ all decay to zero and $w_{\rm eff}\to -0.52$, realizing a late-time accelerating attractor.}
        \label{fig:ecem_results_a}
    \end{figure}

\begin{figure}[htpb]
        \centering
        \subfloat[Phase space for Set (b).]{
            \includegraphics[width=0.72\linewidth]{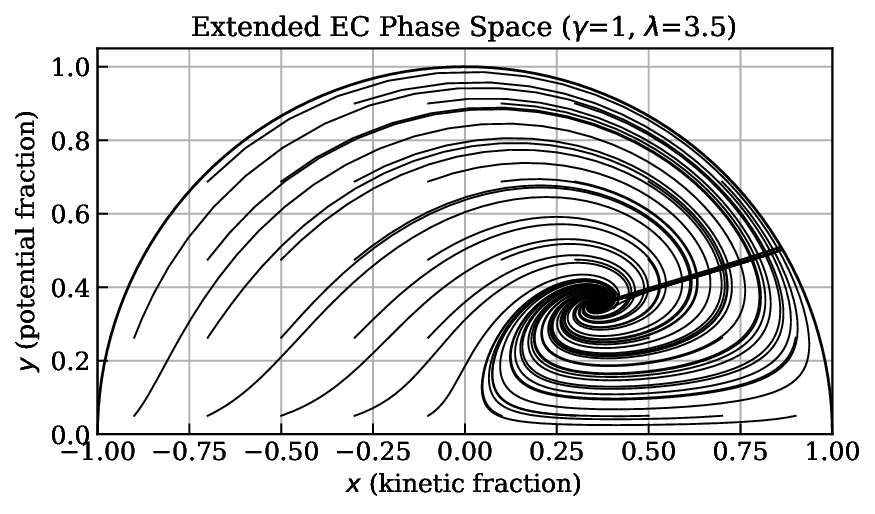}}
        \hfill
        \subfloat[Time evolution for Set (b).]{
            \includegraphics[width=0.72\linewidth]{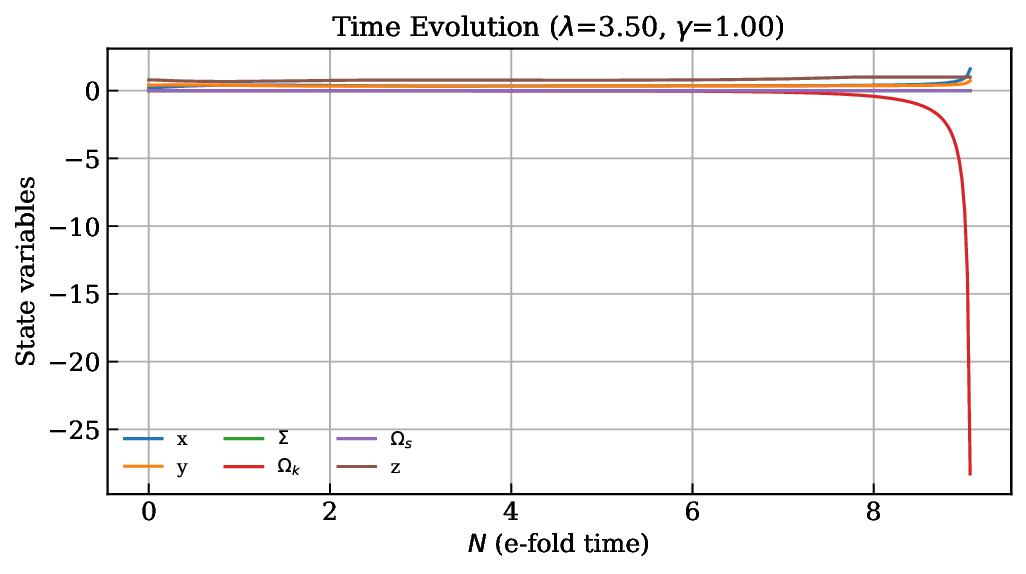}}
        \caption{Set (b), $(\lambda,\gamma_m)=(3.50,1.00)$.  
        (a) Phase portrait in $(x,y)$ for many initial conditions near FLRW. The flow spirals towards the matter-scaling point
        $(x,y)\approx(0.35,0.35)$ with $z\approx0.75$.  
        (b) For a representative trajectory, curvature $\Omega_k$ grows because it redshifts more slowly than matter, so $\Omega_k$ eventually dominates and the system is driven away from the flat scaling regime. In a complete ECEM expand--contract--bounce history (with the full variable $\lambda(\phi)$ potential), one expects a subsequent ekpyrotic contraction (not shown here) to dynamically suppress this curvature growth before any torsion-dominated bounce.}
        \label{fig:ecem_results_b}
    \end{figure}

\begin{figure}[htpb]
        \centering
        \subfloat[Phase space for Set (c).]{
            \includegraphics[width=0.7\linewidth]{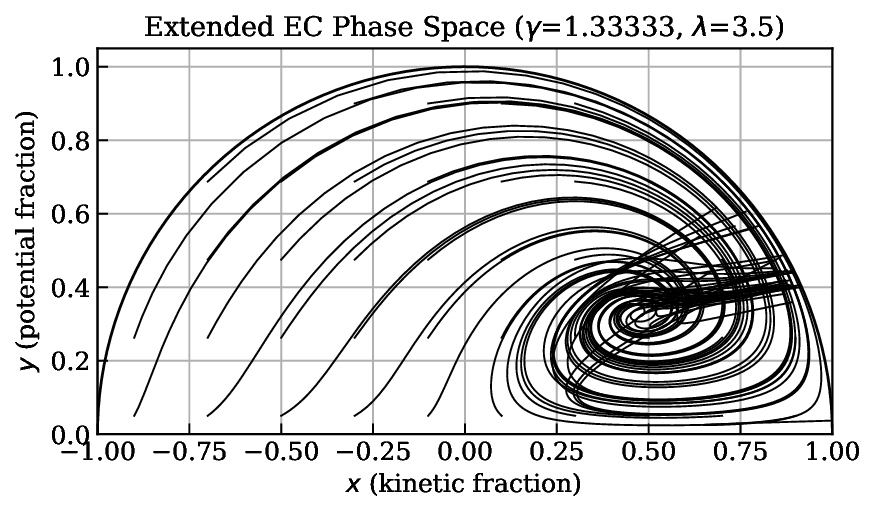}}
        \hfill
        \subfloat[Time evolution for Set (c).]{
            \includegraphics[width=0.7\linewidth]{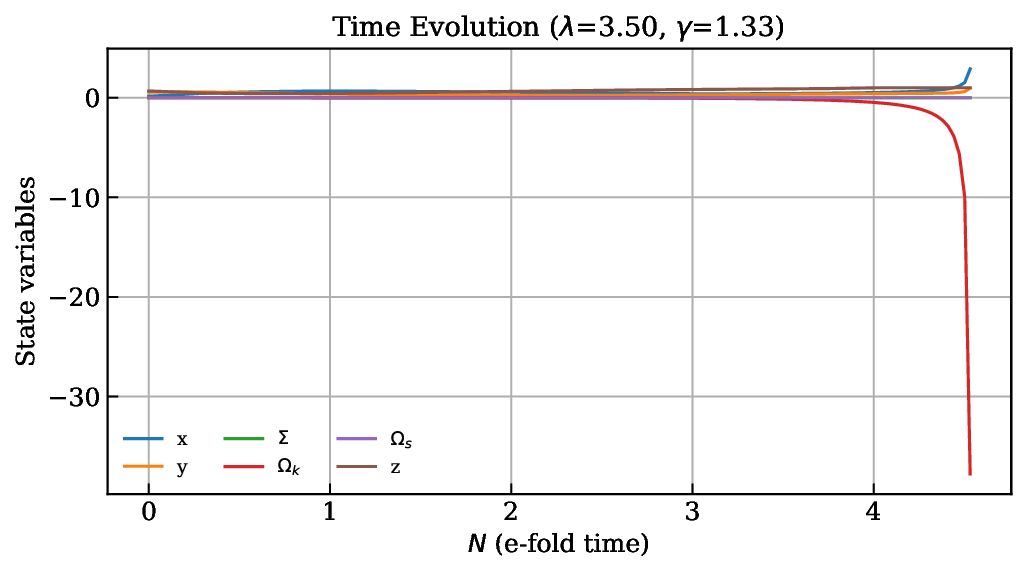}}
        \caption{Set (c), $(\lambda,\gamma_m)=(3.50,4/3)$.  
        (a) Phase portrait in $(x,y)$ showing convergence towards the radiation-scaling point
        $(x,y)\approx(0.47,0.33)$ with $z\approx0.67$.  
        (b) As in the dust case, curvature $\Omega_k$ is unstable at the scaling saddle and eventually dominates at late times, driving the system away from the flat tracking regime. For these parameters no contracting branch or bounce behavior is visible in the $N>0$ evolution.}
        \label{fig:ecem_results_c}
    \end{figure}

\begin{figure}[htpb]
        \centering
        \subfloat[Phase space for Set (d).]{
            \includegraphics[width=0.7\linewidth]{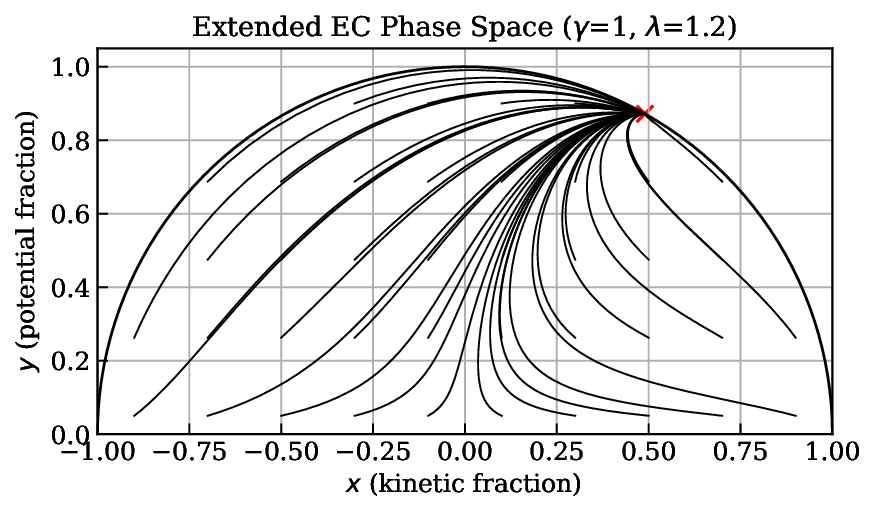}}
        \hfill
        \subfloat[Time evolution for Set (d).]{
            \includegraphics[width=0.7\linewidth]{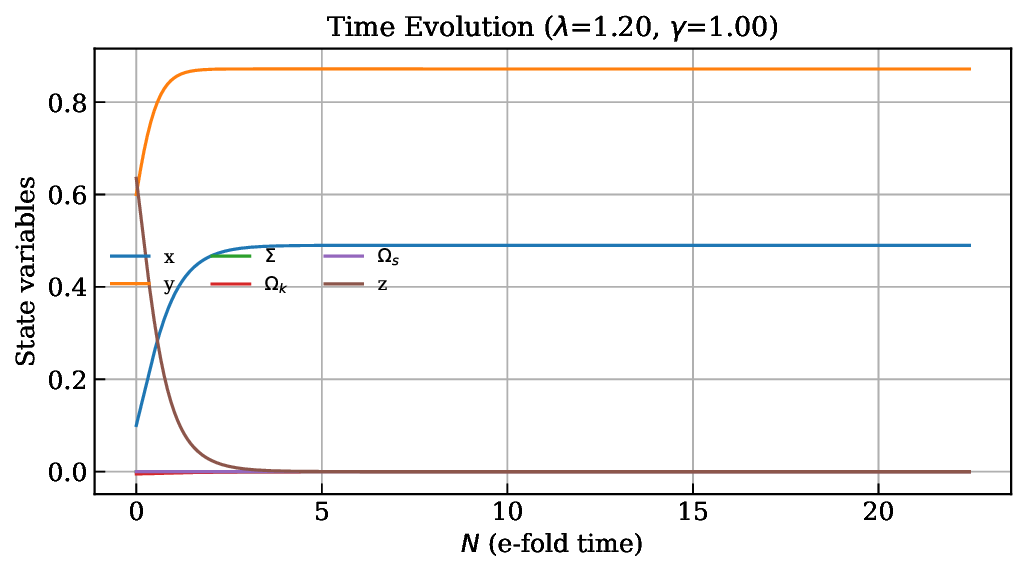}}
        \caption{Set (d), $(\lambda,\gamma_m)=(1.20,1.00)$ with larger initial curvature $\Omega_{k0}=-5\times10^{-3}$.  
        (a) Despite the higher initial $|\Omega_k|$, all trajectories shown converge to the same scalar-field node as in Set (a).  
        (b) For the representative run, curvature, shear, and spin--torsion all decay rapidly, and no curvature-driven deviation or bounce is observed on the expanding branch.}
        \label{fig:ecem_results_d}
    \end{figure}

\begin{figure}[htpb]
        \centering
        \subfloat[Phase space for Set (e).]{
            \includegraphics[width=0.75\linewidth]{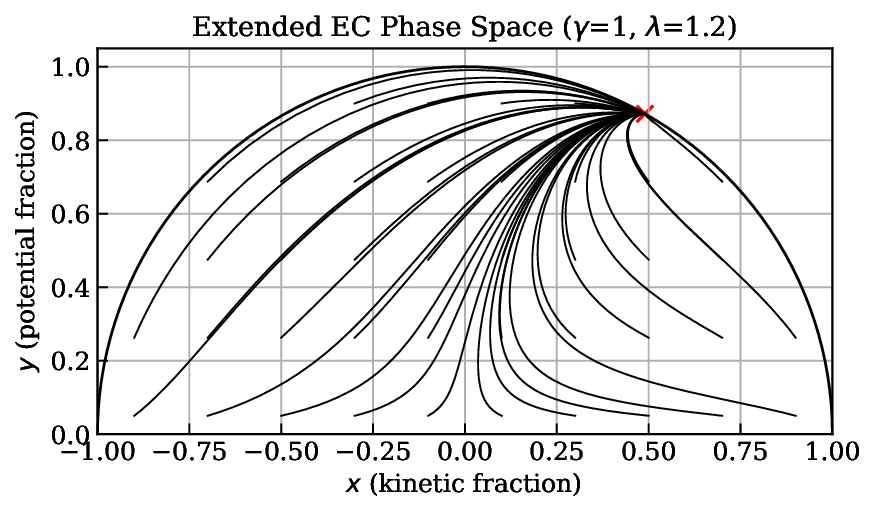}}
        \hfill
        \subfloat[Time evolution for Set (e).]{
            \includegraphics[width=0.75\linewidth]{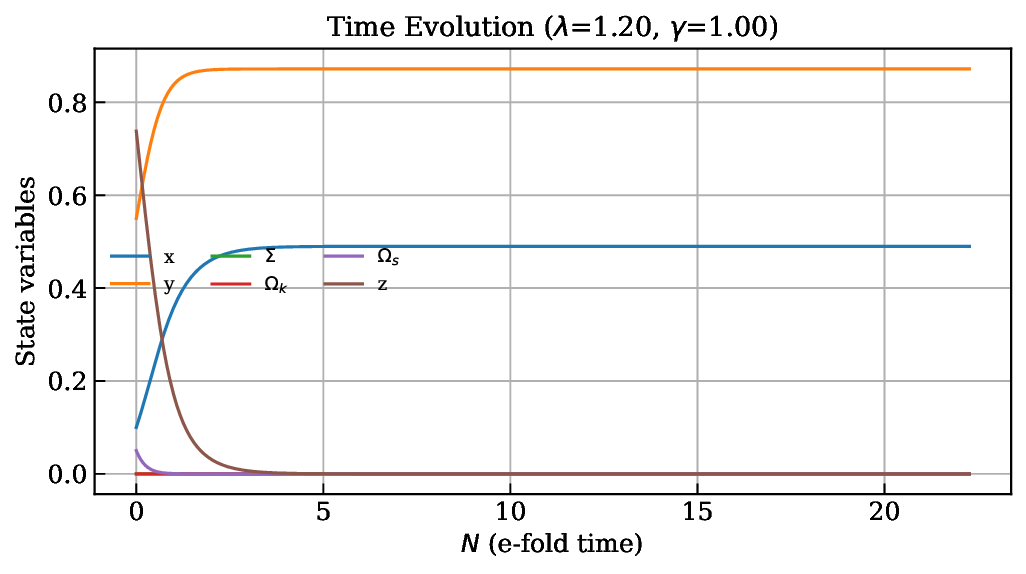}}
        \caption{Set (e), $(\lambda,\gamma_m)=(1.20,1.00)$ with larger initial spin--torsion $\Omega_{s0}=5\times10^{-2}$.  
        (a) The phase portrait remains qualitatively identical to Set (a): trajectories still converge to the scalar-field dominated node.  
        (b) The spin contribution $\Omega_s$ is initially non-zero but decays rapidly (since $\rho_s\propto a^{-6}$), too quickly to produce a visible deviation from the scalar trajectory on this scale. No bounce phase appears in this forward-$N$ evolution.}
        \label{fig:ecem_results_e}
    \end{figure}

    All $N$-time integrations in this appendix use the numerical scheme described in Sec.~\ref{sec:numerics}: the autonomous system \eqref{eq:dyn_xy}-\eqref{eq:dyn_ks} is evolved with fixed $(\lambda,\gamma_m)$, $z(N)$ is reconstructed from the Friedmann constraint \eqref{eq:Friedmann}, and $q(N)$ from Eq.~\eqref{eq:q_canonical}. Here I only summarize the parameter sets and initial conditions used in Figs.~\ref{fig:ecem_results_a}-\ref{fig:ecem_results_e}.

    The initial conditions are varied from run to run. For the parameter sets labelled (a)-(e) in the captions the choices are summarized in Table~\ref{tab:ecem_params}; $z_0$ is always fixed by the constraint. Closed geometry corresponds to $\Omega_k<0$.

    \begin{table}[htpb]
    \centering
    \begin{tabular}{c|c|c|c}
        \hline
        Set & $(\lambda,\gamma_m)$ & $(x_0,y_0)$ & $(\Sigma_0,\Omega_{k0},\Omega_{s0})$ \\
        \hline
        (a) & (1.2, 1.0)   & (0.10, 0.60) & $(10^{-5}, -10^{-4}, 10^{-6})$ \\
        (b) & (3.5, 1.0)   & (0.20, 0.40) & $(10^{-5}, -10^{-4}, 10^{-6})$ \\
        (c) & (3.5, 4/3)   & (0.15, 0.55) & $(10^{-5}, -10^{-4}, 10^{-6})$ \\
        (d) & (1.2, 1.0)   & (0.10, 0.60) & $(10^{-5}, -5\times10^{-3}, 10^{-6})$ \\
        (e) & (1.2, 1.0)   & (0.10, 0.55) & $(10^{-5}, -10^{-4}, 5\times10^{-2})$ \\
        \hline
    \end{tabular}
    \caption{Parameter sets and initial conditions used for the extended Einstein--Cartan runs in Figs.~\ref{fig:ecem_results_a}-\ref{fig:ecem_results_e}. Closed geometry corresponds to $\Omega_k<0$.}
    \label{tab:ecem_params}
    \end{table}
    \FloatBarrier

    \textit{Choice of seeds.}
    The seeds $(\Sigma_0,\Omega_{k0},\Omega_{s0})$ are chosen small enough that the background starts close to FLRW, but deliberately larger than observational bounds to make the geometric dynamics visible in phase space. Since $\rho_s\propto a^{-6}$, even a tiny initial spin fraction grows toward $\mathcal{O}(1)$ as $a$ decreases near any would-be high-density regime. In the runs reported here, changing these seed amplitudes mainly affects visual prominence and transient duration of the geometric sectors, while the fixed-point structure and the curvature-instability trend of the scaling regime remain the same.

    Figure~\ref{fig:ecem_results_a} shows representative time evolution consistent with the flow shown in Fig.~\ref{fig:ecem_intro_phase}(a). The parameters are $(\lambda,\gamma_m)=(1.20,1.00)$ and the initial conditions are Set (a) in Table~\ref{tab:ecem_params}. The orbit converges to the scalar-field dominated node with
    \[
        (x_*,y_*) = \left(\frac{\lambda}{\sqrt{6}},\sqrt{1-\frac{\lambda^2}{6}}\right) \approx (0.49,0.87),
    \]
    and the geometric variables decay,
    \[
        z,\ \Omega_k,\ \Sigma,\ \Omega_s \;\to\; 0.
    \]
    The effective equation of state approaches $w_{\rm eff} \simeq \lambda^2/3-1 \approx -0.52$, corresponding to a late-time accelerating attractor on the expanding branch.

    \FloatBarrier

    Figures~\ref{fig:ecem_results_b}-\ref{fig:ecem_results_e} illustrate how the extended system behaves for steeper potentials and for larger geometric seeds.

    \FloatBarrier

    \FloatBarrier

    \section*{Statements and Declarations}
    \noindent\textbf{Competing interests:} The author has no competing interests.\\
    \noindent\textbf{Funding:} This research received no external funding.\\
    \noindent\textbf{Code availability:} Code developed and used in this study is available from the author upon reasonable request.\\
    \noindent\textbf{Data availability:} This work is purely theoretical; no experimental or observational datasets were generated. Numerical integration outputs underlying the figures are available from the author upon reasonable request.\\
    \noindent\textbf{Use of AI tools:} Large language models were used to assist for copy-editing (English polishing, \LaTeX{} formatting, and paragraph organization). All physics, derivations, code, and scientific conclusions are the author's own, and the author is fully responsible for the content.

    \section*{Acknowledgments}
    The author thanks Dr.~Michal Kopera for encouragement and advice throughout this project, and Dr.~Daryl Macomb for guidance on related high-energy astrophysics research and many helpful discussions. The author is grateful to Boise State University, in particular the Department of Physics for the training that made this work possible and the Department of Mathematics for supporting it as a senior capstone project. The author also thanks his family and friends, especially his parents, for their constant support.

\clearpage
\bibliographystyle{apsrev4-2}
\bibliography{ms}
\end{document}